\documentclass[aps,pre,superscriptaddress,notitlepage,twocolumn,nofootinbib]{revtex4-1}
\usepackage{natbib}
\usepackage{graphicx}
\usepackage{array}
\usepackage[font=footnotesize,justification=raggedright]{caption}
\usepackage{enumerate}
\usepackage{subfig}
\usepackage{float}
\usepackage[hyperindex,breaklinks]{hyperref}
\usepackage{cleveref}
\usepackage{amssymb,amsfonts,amsmath}
\usepackage[export]{adjustbox}
\usepackage{sidecap}
\sidecaptionvpos{figure}{t}
\renewcommand{\thetable}{\arabic{table}}
\renewcommand{\figurename}{Fig. }

\begin{document}

\title{Structure-based control of complex networks with nonlinear dynamics}

\author{Jorge G. T. Za\~nudo}
\email[Corresponding author: ]{jgtz@phys.psu.edu}
\affiliation{Department of Physics, The Pennsylvania State University, University Park, Pennsylvania, 16802-6300, USA.}
\affiliation{Department of Medical Oncology, Dana-Farber Cancer Institute, Boston, Massachusetts 02215, USA.}
\affiliation{Broad Institute of Harvard and Massachusetts Institute of Technology, Cambridge, Massachusetts 02142, USA.}
\author{Gang Yang}
\affiliation{Department of Physics, The Pennsylvania State University, University Park, Pennsylvania, 16802-6300, USA.}
\author{R\'eka Albert}
\affiliation{Department of Physics, The Pennsylvania State University, University Park, Pennsylvania, 16802-6300, USA.}
\affiliation{Department of Biology, The Pennsylvania State University, University Park, Pennsylvania, 16802-5301, USA.}

\begin{abstract}
\textbf{Abstract}: What can we learn about controlling a system solely from its underlying network structure? Here we adapt a recently developed framework for control of networks governed by a broad class of nonlinear dynamics that includes the major dynamic models of biological, technological, and social processes. This feedback-based framework provides realizable node overrides that steer a system towards any of its natural long term dynamic behaviors, regardless of the specific functional forms and system parameters. We use this framework on several real networks, identify the topological characteristics that underlie the predicted node overrides, and compare its predictions to those of structural controllability in control theory. Finally, we demonstrate this framework's applicability in dynamic models of gene regulatory networks and identify nodes whose override is necessary for control in the general case, but not in specific model instances. \\

\textbf{Significance}: Many biological, technological and social systems can be encoded as networks over which nonlinear dynamical processes such as cell signaling, information transmission, or opinion spreading take place. Despite many advances in network science we do not know to what extent the network architecture shapes our ability to control these nonlinear systems. Here we extend a recently developed control framework that addresses this question and apply it to real networks of diverse types. Our results highlight the crucial role of a network's feedback structure in determining robust control strategies, provide a dynamic-detail-independent benchmark for other control methods, and open up a new research direction in the control of complex networks with nonlinear dynamics.
\end{abstract}

\maketitle

Controlling the internal state of complex systems is of fundamental interest and enables applications in biological, technological and social contexts. An informative abstraction of these systems is to represent the system's elements as nodes and their interactions as edges of a network. Often asked questions related to control of a networked system are how difficult to control it is, which network elements need to be controlled, and through which control actions, to drive the system toward a desired control objective \cite{LiuControl,ViczekControl,FVSControl,MotterControl,RuthsControl,MotifControl, GeomControl,RochaControl,Slotine,Sontag,GOptimalControl}. Among control frameworks, structure-based methods distinguish themselves due to their ability to draw dynamical conclusions based solely on network structure and a general assumption about the type of allowed dynamics. E.g., structural controllability, which assumes unspecified linear dynamics or linearized nonlinear dynamics, allows the identification of the minimal number of nodes whose receiving an external signal $u(t)$ drives the system into a state of interest \cite{Lin,Shields}.

Despite its success and wide-spread application \cite{PPIControl,yeastControl,brainControl,SCAkutsu,LiuReview}, structural controllability may give an incomplete view of the network control properties of a system. In case of systems with nonlinear dynamics it provides sufficient conditions to control the system in the neighborhood of a trajectory or a steady state (\cite{LiuControl,LiuReview}, SI Appendix), and its definition of control (full control; from any initial to any final state) does not always match the meaning of control in biological, technological, and social systems, in which control tends to involve only naturally occurring system states \cite{Muller}. In addition to the approaches provided by nonlinear control theory \cite{Slotine,Sontag,GOptimalControl,LiuReview}, new methods of network control have been proposed to incorporate the inherent nonlinear dynamics of real systems and relax the definition of full control \cite{MotterControl,MotifControl,GOptimalControl,MurragarraControl,LiuReview}. Only one of these methods, namely feedback vertex set control (FC), can be reliably applied to large complex networks in which only the structure is well known and the functional form of the governing equations is not specified. This method, introduced by Fiedler, Mochizuki et al in \cite{FVSControl,FVSMath}, incorporates the nonlinearity of the dynamics and considers only the naturally occurring end states of the system (e.g. steady states and limit cycles) as desirable final states. In this work, we use feedback vertex set control on biological, technological, and social networks to predict the nodes whose override (by external control) can steer a network's dynamics towards any of its natural long term dynamic behaviors (its dynamical attractors). We identify the topological characteristics underlying the predicted node overrides, compare the obtained results with those of control theory's structural controllability \cite{LiuControl,Lin,Shields} and identify the model-dependent and model-independent overrides it provides for network models with parameterized dynamics.

\section*{Structure-based network control with nonlinear dynamics} \label{sec:StrCtrl}

Most real systems are driven by nonlinear dynamics in which a decay term prevents the system's variables from increasing without bounds. The state of the system's $N$ nodes at time $t$, characterized by source node variables $S_j(t)$ (for nodes with no incoming edges) and internal node variables $X_i(t)$, obeys the equations
\noindent\begin{minipage}{.5\linewidth}
\begin{align}
dX_i/dt = F_i(X_i,X_{I_i},t), \label{eq:1}
\end{align}
\end{minipage}%
\begin{minipage}{.5\linewidth}
\begin{align}
dS_j/dt = G_j(t), \label{eq:2}
\end{align}
\end{minipage}
where $i=1, \ldots, N-N_s$, $j=N-N_s+1, \ldots, N$, and $N_s$ is the number of source nodes. The dynamics of each source node $j$ is independent of the internal node variables $X_i$ (by definition), is fully determined by $G_j(t)$, and does not include a decay term. In the simplest case $G_j=0$ and $S_j$ will remain in its specified initial value. The dynamics of each internal node $i$ is governed by $F_i(X_i,X_{I_i},t)$, which captures the nonlinear response of node $i$ to its predecessor nodes $I_i$ (which can be source or internal nodes), and which includes decay in the dependence of $F_i$ on $X_i$ (\hyperref[sec:SText]{SI Appendix}). Functions of the form $F_i=f_i(X_{I_i})-\alpha_i(X_{I_i}) X_i$, which satisfy these conditions, are used to describe the dynamics of birth-death processes \cite{StochProc,BirthDeath}, epidemic processes \cite{StochProc,Epidemic,SocialDynamics}, biochemical dynamics \cite{TysonReview,AlonBook}, and gene regulation \cite{TysonReview,AlonBook,SegPolODE,SegPolBool}. As an example, $X_i(t)$ can denote the concentration of proteins involved in a signal transduction pathway, and $S_j(t)$ the concentration of extracellular signals (molecules). In this case $f_i$ can take the form of a Hill function (e.g. $f_i=\beta_i X_k^2/(X_k^2 + \theta^2)$ if $k$ is the only node in $I_i$) or of a mass-action term (e.g. $f_i=\beta_i X_k X_l$ if $k$ and $l$ are the only nodes in $I_i$). As an alternative example, $X_i(t)$ can denote the probability that an individual is infected in a contagion network and $S_j(t)$ the influence of vaccination or prevention measures on certain individuals, and $F_i$ can take the form of a susceptible-infected-susceptible model term (e.g. $F_i=\beta_i X_k(1-X_i)-\alpha_i X_i$ if $k$ is the only node in $I_{i}$).\footnote{ Note that these functions are just examples, and that the framework we describe is valid for any bounded dynamical process of the form of Eqs. \ref{eq:1}-\ref{eq:2} that occurs on the specified network structure.}

The dynamics described by Eqs. \ref{eq:1}-\ref{eq:2} are such that they possess some naturally occurring end states, or dynamical attractors. Dynamical attractors in biological, social, and technological systems represented by networks have been found to be identifiable with the stable patterns of activity of the system. E.g., in gene regulatory networks dynamical attractors correspond to cell fates \cite{AlonBook,SegPolODE,SegPolBool}; in opinion spreading dynamics on social networks they correspond to opinion consensus states of groups of individuals \cite{SocialDynamics}; and in disease or computer virus spreading they correspond to the long-term (endemic) patterns of infected elements \cite{Epidemic}.

In many systems there is adequate knowledge of the underlying wiring diagram but not of the specific functional forms and parameter values required to fully specify $F_i$ and $G_j$. Analyzing such systems requires the use of structure-based control methods such as feedback vertex set control (FC). FC, developed by Fiedler, Mochizuki et al. \cite{FVSControl,FVSMath}, is a mathematical formalization of the following idea: in order to drive the state of a network to any one of its naturally occurring end states (dynamical attractors) one needs to manipulate a set of nodes that intersects every feedback loop in the network - the feedback vertex set (FVS). This requirement encodes the importance of feedback loops in determining the dynamical attractors of the network, a fact that was recognized early on in the study of the dynamics of biological networks \cite{Thomas,Kauffman}. Fiedler, Mochizuki et al. mathematically proved that for a network governed by the nonlinear dynamics of Eq. \ref{eq:1}, the control action of forcing (overriding) the state variables of the FVS into the trajectory specified by a given dynamical attractor of Eq. \ref{eq:1} ensures that the network will asymptotically approach the desired dynamical attractor, regardless of the specific form of the functions $F_i$. Note that FC does not utilize a controller or driver signal, and instead considers node state override as its control action\footnote{The general task of designing a controller with an attractor as the target state in a nonlinear system is a difficult and unsolved problem that depends strongly on the functions $F_i$ although several numerical algorithms for specific types of controllers have been proposed (\cite{MotterControl,GeomControl,LiuReview}, \hyperref[sec:SText]{SI Appendix}).}. This type of intervention is often used in biological systems, with examples such as genome editing or pharmacological treatment \cite{Muller,BioControl1}, and in epidemic spreading networks, where vaccination is a node state override that prevents a node from being infected. When using node state overrides as the control action, controlling the FVS is sufficient to drive the system to any of its attractors for each form of $F_i$ and necessary if this must hold for every $F_i$ (\cite{FVSControl,FVSMath} and \hyperref[sec:SText]{SI Appendix}). The problem of exactly identifying the minimal FVS is NP-hard, but a variety of fast algorithms exist to find close-to-minimal solutions (SI Appendix).

\begin{figure}
\centering
\includegraphics{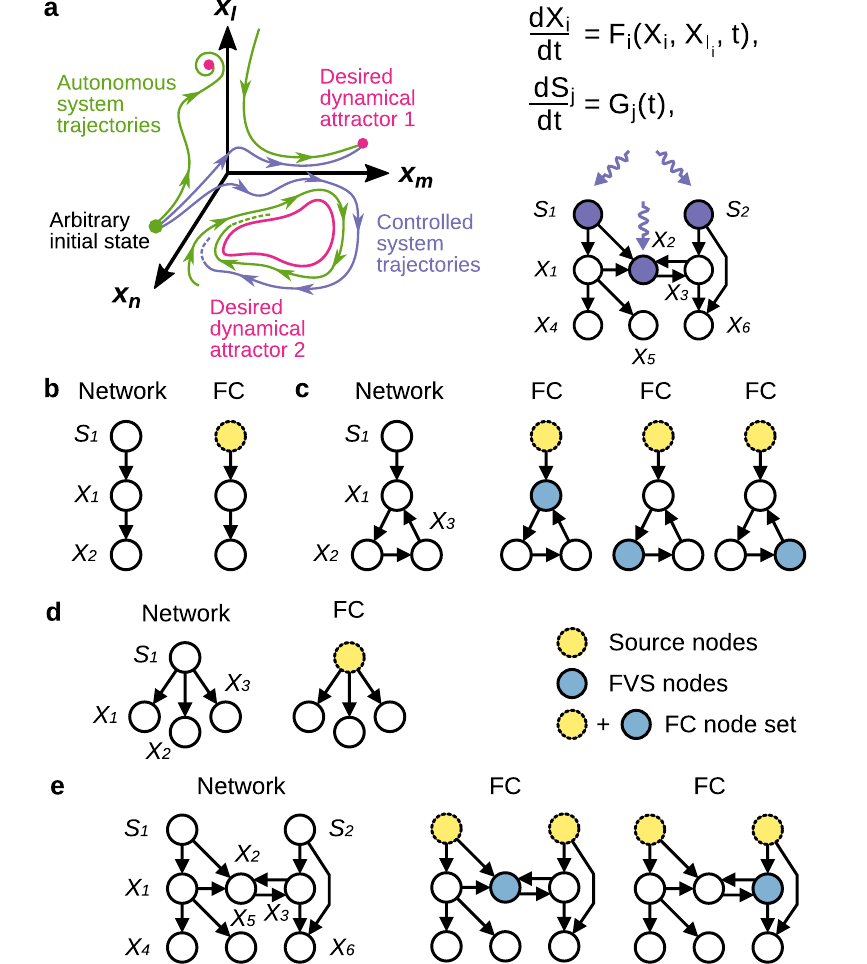}
\caption{Structure-based network control with nonlinear dynamics. Feedback vertex set control (FC) is a structure-based control method that can make conclusions about the long-term dynamics of a system using solely the network structure. (a) In FC the objective is to drive the network from an arbitrary initial state to any desired dynamical attractor of the system (e.g. a steady state) by forcing (overriding) the state variables of certain nodes. (c-f) FC in simple networks. FC requires control of the source nodes (yellow nodes with dotted outlines) and of all cycles by control of the feedback vertex set (FVS, blue nodes with solid outlines).}
\label{fig:ControlFigure}
\label{fig:ControlExample}
\end{figure}

In the structural theory of Fiedler, Mochizuki et al. every element is governed by Eq. \ref{eq:1}. It is assumed that the source nodes converge to a unique state (or trajectory) and do not need independent control; thus they are iteratively removed from the network prior to applying FVS control. However, source nodes can denote external stimuli or boundary conditions the system is subject to; a different set of attractors may be available for each state of a source node. E.g., in the parameterized biological models we consider, source nodes provide positional information for the cells and affect the patterning behaviors cells are capable of.

\begin{figure*}
\centering
\includegraphics{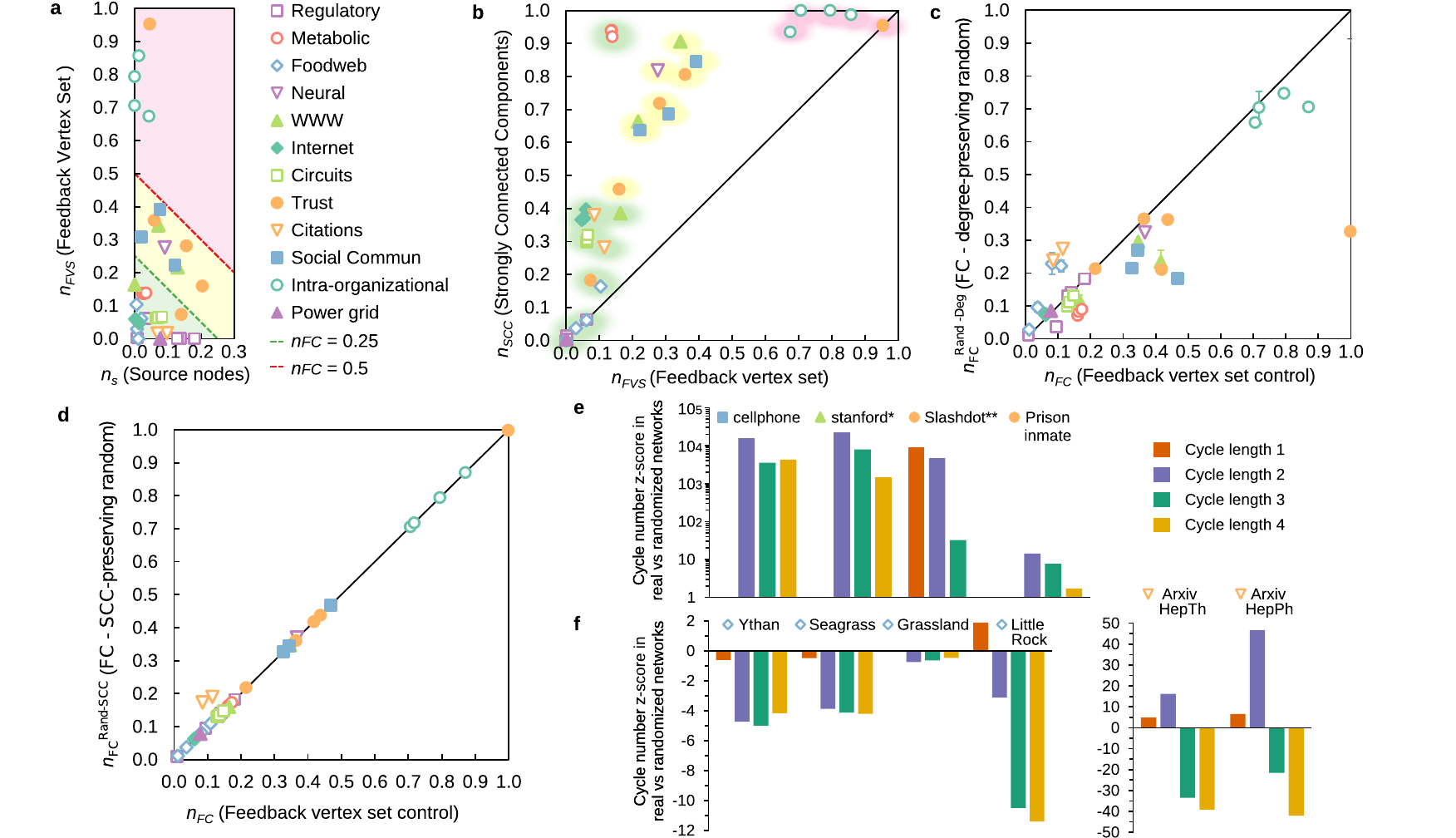}
\caption{Feedback vertex set control in real networks. (a) Scatter plot with the contribution of source nodes $n_s$ and the feedback vertex set $n_{FVS}$ to the fraction of control nodes in feedback vertex set control $n_{FC}$ for each real network in SI Appendix, Table S1. Each of the two lines in the scatter plot correspond to a fixed value of $n_{FC}$ , since $n_{FC}= n_{s}+ n_{FVS}$. The background color of the scatter plot indicates areas in which $n_{FC}$ takes a certain range of values: green for $n_{FC}<0.25$, yellow for $0.25<n_{FC}<0.5$, and pink for $n_{FC}>0.5$. (b) Scatter plot of the fraction of nodes in the feedback vertex set (FVS) $n_{FVS}$ and the fraction of nodes in a strongly connected component (SCC) $n_{SCC}$ for each real network. The shading of the symbols corresponds to their position in panel a and reflects the relative size of their FC node set. (c, d) Scatter plot with the fraction of control nodes in FC for real networks ($n_{FC}$) and their degree-preserving randomization ($n_{FC}^{Rand-Deg}$) (panel c) or SCC-preserving randomization ($n_{FC}^{Rand-SCC}$) (panel d). Error bars denote the estimated standard deviation of the randomized ensembles. (e, f) Cycle number z-score for different cycle lengths in real versus degree-preserving randomized networks for the networks with $n_{FC}>>n_{FC}^{Rand-Deg}$ (panel e) and $n_{FC}<< n_{FC}^{Rand-Deg}$ (panel f).}
\label{fig:FVSRandom}
\end{figure*}

Here we adapt the structural theory of Fiedler, Mochizuki et al. to networks in which source nodes are governed by Eq. \ref{eq:2} (Fig. \ref{fig:ControlFigure}a and \hyperref[sec:SText]{SI Appendix}). Since the source nodes are unaffected by other nodes, one additionally needs to lock the source nodes of the network in the trajectory specified by the attractor. We emphasize that the treatment of source nodes is not merely cosmetic, since the state of a source node can affect the dynamical attractors available to the system. E.g., steady states can merge, appear, or disappear depending on the presence or absence of an external stimulus represented by a source node \cite{TysonReview,SocialBassett}. In summary, control of the source nodes and of the FVS of a network guarantees that we can guide it from any initial state to any of its dynamical attractors (i.e., its natural long term dynamic behaviors) regardless of the specific form of the functions. In the following we refer to this attractor-based control method as feedback vertex set control (FC) (Fig. \ref{fig:ControlFigure}a), and to the group of nodes that need be manipulated FC as a FC node set.

To illustrate FC, consider the example networks in Fig. \ref{fig:ControlExample}. In a linear chain of nodes (Fig. \ref{fig:ControlExample}b, left) the only node that needs to be controlled is the source node $S_1$. For Fig. \ref{fig:ControlExample}c, a source node connected to a cycle, FC requires controlling the source node $S_1$ and any node $X_i$ in the cycle, the FVS in this network. Fig \ref{fig:ControlExample}d consists of a source node with three successor nodes, and FC requires controlling only the source node $S_1$ since there are no cycles in the network. In Fig \ref{fig:ControlExample}e we show a more complicated network with a cycle and several source and sink nodes, and two minimal FC node sets. These examples illustrate an important feature of FC, namely, that it is determined by the cycle structure and the input layer of the network. SI Appendix, Fig. S1 illustrates FC in a network in which a specific form of the functions $F_i$ and $G_j$ is given.

\section*{Feedback vertex set control of real networks} \label{sec:RealNetworks}

We applied FC to several real networks and the ratio of the minimal FC node set, $N_{FC}$, and the total number of nodes, $n_{FC}=N_{FC}/N$ was used to gauge how difficult it is to control these networks. The real networks are of diverse types (biological, technological, and social) and various sizes (from dozens to millions of elements), and have been repeatedly used as benchmarks to study structural controllability (SC) \cite{LiuControl,LiuReview}. The FC results are shown in SI Appendix, Table S1 and Fig. \ref{fig:FVSRandom}a, where the feedback vertex set and source node contributions of $n_{FC}$ are denoted by $n_{FVS}$ and $n_{s}$, respectively ($n_{FC}=n_{FVS}+n_{s}$). We observed that most types of biological networks (gene regulatory, metabolic, and food web networks) require control of a smaller fraction of nodes than social networks (trust, social communication, and intra-organizational networks); $n_{FC}$ is between 1\% - 18\% in biological networks vs. more than 21\% in social networks. FC's prediction that biological networks are easier to control than social networks matches recent experimental results in cellular reprogramming and large-scale social network experiments \cite{Muller,SocialBassett,SocialFacebook}.

\begin{figure}
\centering
\includegraphics{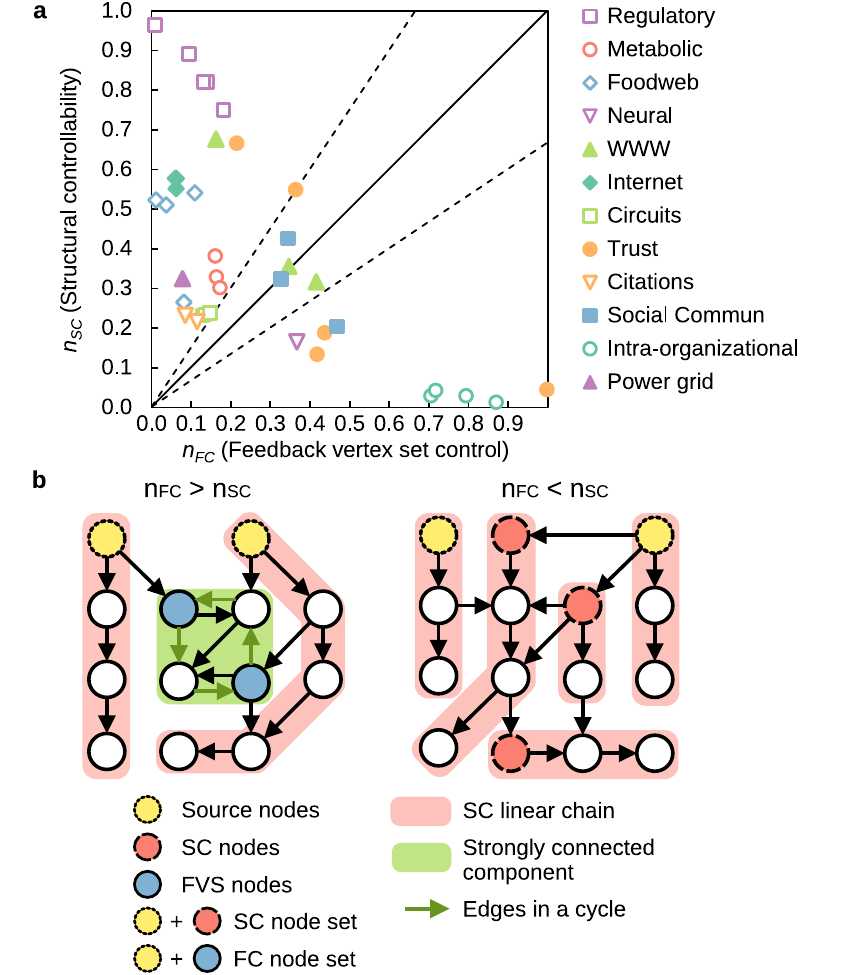}
\caption{Comparing feedback vertex set control (FC) and structural controllability (SC). (a) Scatter plot with the fraction of control nodes in FC ($n_{FC}$) and SC ($n_{SC}$) for each real network in SI Appendix, Table S1. The bold line denotes the positions in the plot with $n_{SC}=n_{FC}$, while the dashed lines denote $n_{SC}=1.5 \ n_{FC}$ and $n_{FC}=1.5 \ n_{SC}$. (b) Examples of the effect of cycle structure in the FC and SC node set size. Control of the source nodes (yellow nodes with dotted outlines) is shared by SC and FC; in SC every source node is the top node of a chain in a minimal group of non-intersecting linear chains of nodes (pink background) and directed cycles (green edges) that span the network. SC additionally requires controlling the top nodes in other chains (red nodes with dashed outlines) but requires no independent control of cycles. FC requires controlling all cycles by control of the feedback vertex set (blue nodes with solid outlines).}
\label{fig:FVSvsSC}
\end{figure}

To understand the topological properties underlying the diversity of the fraction of control nodes $n_{FC}$ among networks, we identify the nodes in a network that determine its cycle structure, and thus, the FVS contribution of the control nodes ($n_{FVS}$). Every node that is involved in a cycle must also be part of a strongly connected component (SCC), a group of nodes in a network in which there is a directed path between any pair of nodes. The concept of SCC is related to the bow-tie structure of multiple empirical directed networks \cite{NewmanBook}, in which most of the network belongs to a large SCC, its in-component (the nodes that can reach the SCC), or its out-component (the nodes that can be reached from the SCC).

Applying this reasoning to the studied real networks (SI Appendix, Table S1), we expect the networks in which the fraction of nodes that are part of an SCC is high to have a large FVS contribution $n_{FVS}$. As shown in Fig. \ref{fig:FVSRandom}b, the networks show a strong correlation between the relative size of their SCCs (denoted by $n_{SCC}$) and of their FVS (SI Appendix, Fig. S2a). For example, all of the networks with the largest FC node set size ($n_{FC}>0.5$, Fig. \ref{fig:FVSRandom}a,b, pink shading; e.g. intra-organizational networks) have a large fraction of nodes in their SCCs ($n_{SCC}>0.93$). Similarly, networks with an intermediate FC node set size ($0.25<n_{FC}<0.5$, Fig. \ref{fig:FVSRandom}a,b, yellow shading; e.g. social communication networks, and most trust and WWW networks) have an intermediate $n_{SCC}$ ($0.46<n_{SCC}<0.91$), and most of the networks with the smallest FC node set size ($n_{FC}<0.25$, Fig. \ref{fig:FVSRandom}a,b, green shading; e.g. food webs, circuits, and gene regulatory networks) have correspondingly small SCCs ($n_{SCC}<0.4$).

Motivated by the observed remarkable agreement between the number of control nodes of real networks and their degree-preserving randomized versions in SC \cite{DegRandomized,LiuControl}, we study FC in similarly randomized networks (SI Appendix, Table S1 and \hyperref[sec:SText]{SI Text}). We find much weaker agreement: for most networks the number of FC nodes is higher than the number of control nodes in randomized versions ($n_{FC}>n_{FC}^{Rand-Deg}$), with the notable exceptions of food web and citation networks, in which randomized networks require more control nodes ($n_{FC}<n_{FC}^{Rand-Deg}$), (Fig. \ref{fig:FVSRandom}e,f). A closer look reveals that the cycle structure of the real networks - their cycles and SCCs - is responsible for the discrepancy of $n_{FC}$. Although the number of nodes in a SCC is similar or smaller compared to their degree-preserving randomized counterparts, real networks tend to have a more complicated cycle structure, evidenced by the over-representation of short cycles compared to the randomized networks (Fig. \ref{fig:FVSRandom}e), and reflected by the larger size of their FVS (SI Appendix, Table S1). The exception to this reasoning are food web and citation networks (Fig. \ref{fig:FVSRandom}f), which are known to have an acyclic (e.g. tree-like) or close-to-acyclic structure \cite{NewmanDAG}, and thus, feature fewer cycles and fewer nodes in a SCC than randomized networks.

To verify that the cycle structure of real networks explains the observed FC node set size, we generated degree preserving randomized versions of these networks that maintain their cycle structure, which we achieve by randomizing the directed acyclic part of the graph while keeping intact the SCCs (\hyperref[sec:SText]{SI Appendix}). The results show a remarkable agreement between the FC node set size of the networks and their randomized versions (Fig. \ref{fig:FVSRandom}d and SI Appendix, Table S1). Given that short cycles were found to correlate well with the discrepancy in FC node set size in real networks compared to randomized networks (SI Appendix, Table S1, Fig. \ref{fig:FVSRandom}e,f), we reasoned that preserving only the short-cycle structure of networks (in addition to their degree) might be sufficient to explain the FC node set size of real networks. To test this, we generated degree-preserving randomized versions of the networks that maintain their short-cycle structure (cycles of length 4 or less) (\hyperref[sec:SText]{SI Appendix}). SI Appendix, Fig. S2b and Table S1 show the resulting FC node set sizes, which have an excellent agreement with that of the real networks, the exceptions being the near-acyclic food web and citation networks, for which short cycles cannot capture their near-acyclic structure.

Taken together, these results show that the cycle structure of a network, specifically its SCCs and short cycles, determines the number of nodes that need to be overridden in FC.

\begin{figure}
\centering
\includegraphics{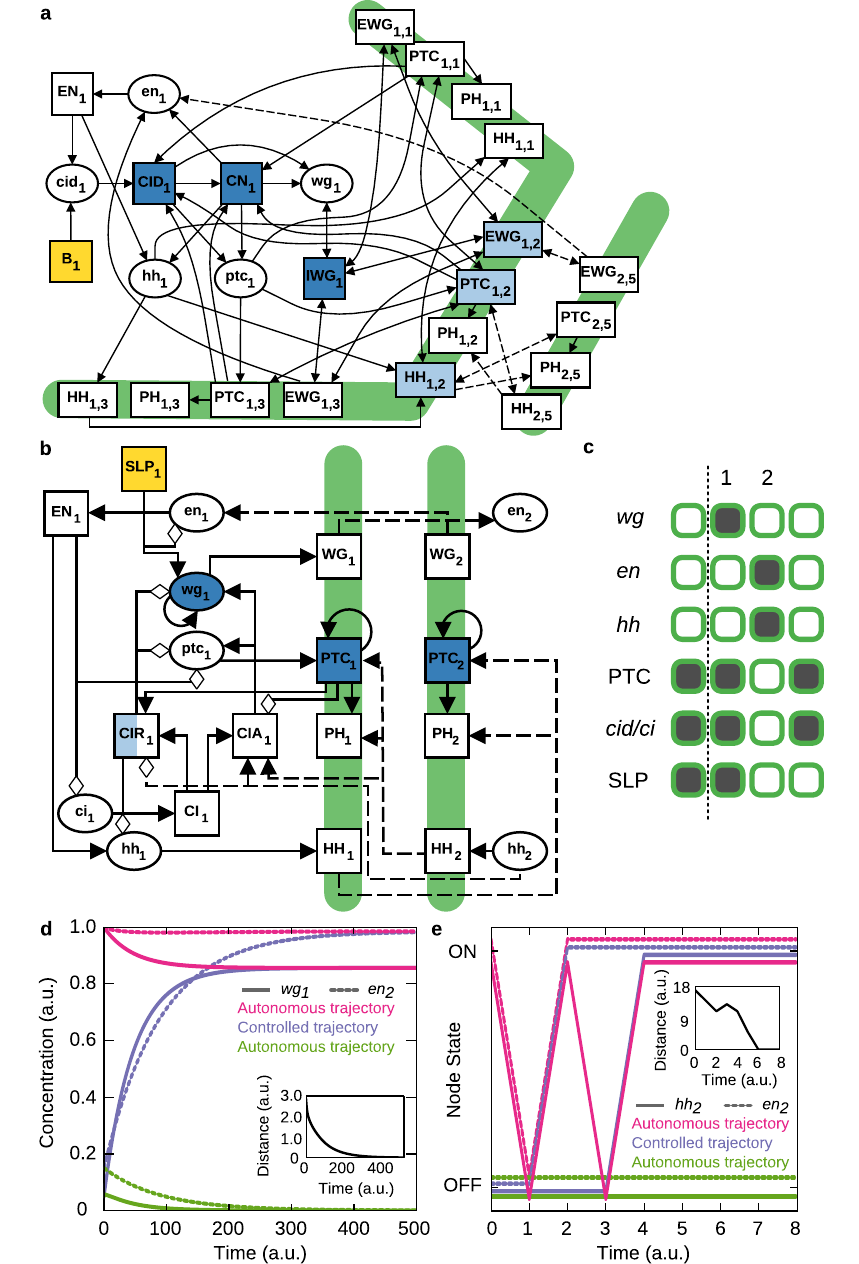}
\caption{Control of the Drosophila segment polarity network models. (a, b) Networks corresponding to the differential equation model (panel a) and the discrete model (panel b). Each figure shows one cell of the four-cell parasegment together with the cell boundaries (thick green lines); the complete networks contain four cells in a symmetric completion of each figure. Elliptical nodes denote mRNAs and rectangular nodes denote proteins, which can be localized inside the cell or in the membrane (subscripts refer to the cell number and surface index). Intracellular interactions are drawn with solid lines and intercellular interactions are dashed. In panel b, positive edges are drawn with black arrowheads and negative edges with white diamonds. Yellow nodes are source nodes, blue nodes are FC nodes in every cell, and half white/half blue nodes are FC nodes in alternating cells. Dark blue nodes are sufficient for attractor control in the considered dynamic models. (c) Wild type segment polarity gene product expression pattern in a Drosphila parasegment. The parasegment boundary (dotted line) is between the \textit{wg}-expressing cells (cell 1) and \textit{en}-expressing cells (cell 2). (d, e) The dynamics of \textit{wg} in the first cell (panel d, solid lines) and \textit{hh} in the second cell (panel e, solid lines), and \textit{en} in the second cell (dotted lines) in the models. Pink lines and green lines represent autonomous trajectories that start from different initial conditions and converge to different steady states (the wild type state and the unpatterned state, respectively). Blues lines represent the case when the system starts from the initial condition that autonomously evolves to the unpatterned state, but when applying FC, evolves into the wild type steady state. Insets: evolution of the norm of the difference between the desired attractor and the controlled state trajectory using FC.}
\label{fig:DynMod}
\end{figure}

\section*{Comparing feedback vertex set control and structural controllability}

An interesting result from applying FC on real networks is that biological networks are easier to control than social networks, yet this prediction stands in contrast with those of structural controllability (SC) on the same type of networks, in which the opposite result was obtained \cite{LiuControl}. This contradicting prediction is somewhat surprising, since both methods can be used to answer the question of how difficult to control a network is based solely on network structure, albeit each focuses on a different aspect of control (full control vs. attractor control), considers different underlying dynamics (linear vs. nonlinear), and uses different control actions (controller signal vs. node state override). To test whether this significant difference in the predictions of FC and SC is common among other networks, we compare their fraction of control nodes $n_{SC}$ and $n_{FC}$. As shown on Fig. \ref{fig:FVSvsSC}a and SI Appendix, Table S1, $n_{SC}$ and $n_{FC}$ appear to be inversely related across several types of networks. E.g., gene regulatory networks require between 75\% - 96\% of nodes in SC yet only require between 1\% - 18\% of nodes in FC. A similar $n_{SC}>>n_{FC}$ relationship is also seen in food web networks and internet networks, while the opposite relationship ($n_{SC}<<n_{FC}$) is seen in the social trust networks with low $n_{SC}$ and intra-organizational networks. This difference between methods warns practitioners against a naive application of SC or FC to control situations beyond their realm of applicability in terms of dynamics, control objective, or control action, as others have previously cautioned \cite{RochaControl}.

The difference in $n_{SC}$ and $n_{FC}$ can be attributed to the treatment of cycles in each of these methods: cycles have to be controlled in FC but do not require independent control in SC. In SC, the nodes that must be directly controlled are each node at the top of a (minimal) group of non-intersecting linear chains of nodes and directed cycles that span the network; these cycles do not need to be directly controlled if there is a path to them from a linear chain of nodes.

To illustrate how the cycle structure influences the number of control nodes in FC and SC, consider the networks in Fig. \ref{fig:FVSvsSC}b. The left-most network contains several cycles (green background) and requires more nodes to be manipulated in FC compared to SC ($n_{FC}>n_{SC}$). In FC each of these cycles can be controlled through the nodes in the FVS (blue nodes); in SC, the cycles do not require independent control given that the whole network is spanned by the specified group of linear chains of nodes (pink background) and a directed cycle (green edges). The right-most network in Fig. \ref{fig:FVSvsSC}b has $n_{FC}<n_{SC}$ because of the absence of cycles, which means FC only requires controlling the source nodes (yellow nodes) while SC requires additional nodes (red nodes) because of the group of non-intersecting linear chains. A detailed analysis in which the topological properties underlying SC and FC are jointly considered backs up the importance of the cycle structure in the differences between their results and points to other contributing factors (\hyperref[sec:SText]{SI Appendix}).

\section*{Feedback vertex set control and dynamic models of real systems} \label{sec:DynModels}

Validated dynamic models can be an excellent testing ground to assess control methods \cite{MotterControl,MotifControl,RochaControl}. Here we use two models for the gene regulatory network underlying the segmentation of the fruit fly (\textit{Drosophila melanogaster}) during embryonic development: a differential equation (ODE) model by von Dassow et al. \cite{SegPolODE} (Fig. \ref{fig:DynMod}a) and a discrete (Boolean) model by Albert and Othmer \cite{SegPolBool} (Fig. \ref{fig:DynMod}b). Both models consider a group of four subsequent cells as a repeating unit, include intracellular and intercellular interactions among proteins and mRNAs, and both recapitulate the observed (wild type) stable pattern of gene expression (Fig. \ref{fig:DynMod}a-c and \hyperref[sec:SText]{SI Appendix}).

Using FC on these network models, we find $N_{FC}=52$ ($14$) for the ODE (discrete) model (Fig. \ref{fig:DynMod}a-c, SI Appendix, Fig. S3, and \hyperref[sec:SText]{SI Appendix}). Both model networks have a large SCC, and thus, a significant FVS contribution to the FC node set; $n_{SCC}$/$n_{FVS}$ are 0.74/0.35 and 0.5/0.18, respectively, similarly to the yellow-shaded networks in Fig. \ref{fig:FVSRandom}. In FC, locking the FC nodes into their trajectory in the wild type attractor successfully steers the system to the wild type attractor (Fig. \ref{fig:DynMod}d-e and SI Appendix, Figs. S4-S5 and \hyperref[sec:SText]{SI Text}). Thus, FC gives a control intervention that is directly applicable to dynamic models and that is directly linked to their long-term behavior.

FC gives a sufficiency condition about the ensemble of all models with a given network structure, and consequently, a subset of the FC node set can often be sufficient for a given model and an attractor of interest (i.e. FC provides an upper limit for the size of the control node set). For the fruit fly gene regulatory models we show that 16 (12) nodes are sufficient for the continuous (discrete) model, respectively, which is a 66\% (14\%) reduction (Fig. \ref{fig:DynMod}a-c and SI Appendix, Figs. S4-S5 and \hyperref[sec:SText]{SI Text}). Similar results were obtained in \cite{FVSMath}, who found that 5 nodes (out of 7 in the FVS) are sufficient for attractor-based control in a model of the mammalian circadian rhythm. The generality of these findings is supported by a recently developed control method in which controlling a subset of the cycles (and, thus, a subset of the FVS) in Boolean dynamic models was proven to be sufficient for attractor control (\cite{MotifControl}, \hyperref[sec:SText]{SI Appendix}). This shows that FC provides a benchmark of attractor control node sets that are model independent, as well as an upper limit to model dependent control sets.

\section*{Discussion}

Network control methods have the general objective of identifying network elements that can drive a system toward a specified goal while satisfying a set of constraints. Different control methods answer complementary aspects of control in a complex network; which one to use depends on the specific question being asked, on the natural definition of control and the underlying dynamics in the system or discipline of interest. We argue that attractor-based control (and, thus, FC) is the appropriate choice of control for biological systems, for which a long history of dynamic modeling has established the correspondence of attractors with biological states of interest \cite{AlonBook}, but also in many social and technological contexts, as illustrated by opinion dynamics and the consensus state, and by epidemic processes and the endemic state \cite{SocialDynamics,Epidemic}.

As we showed in this work, FC is directly applicable to systems in which only structural information is known, and also to systems in which a parameterized dynamic model is available, for which it provides realizable control strategies that are robust to changes in the parameters and functions. FC also provides a benchmark and a point of contact with the large body of work in network control methods that require the network structure and a dynamic model \cite{MotterControl,MotifControl,MurragarraControl,LiuReview,RochaControl}. The prescription of a directly realizable control action (even if a controller signal is not provided) has no analogue in control theory's structure-based methods such as structural controllability, wherein the existence of a controller signal is guaranteed but it is yet to be determined. SC instead has the advantage of integrating controller signals into its framework, and being a well-developed concept in control and systems theory with connections to other notions of control in linear and nonlinear systems \cite{Slotine,Sontag,GOptimalControl}. Further work is needed to extend FC and address questions such as the level of control provided by a subset of nodes, the task of building a controller signal that can implement the node state overrides, and the difficulty of steering the system towards a desired state, concepts which are well-developed in control theory \cite{Slotine,Sontag,GOptimalControl,LiuReview}. Taken together, our work opens up a new research direction in the control of complex networks with nonlinear dynamics, connects the field of dynamic modeling with structure-based methodologies, and has promising theoretical and practical applications.

\begin{acknowledgments}
We thank A Mochizuki and MT Angulo for helpful discussions, and YY Liu for his assistance and for providing us some of the networks in this study. We also thank the MBI for the workshop ``Control and Observability of Network Dynamics'', which greatly enriched this paper. This work was supported by NSF grants PHY 1205840, 1545832, and IIS 1160995. JGTZ is a recipient of a SU2C - V Foundation Convergence Scholar Award.
\end{acknowledgments}

\renewcommand\theequation{S\arabic{equation}}
\renewcommand{\tablename}{}
\renewcommand{\thetable}{Table S\arabic{table}}
\renewcommand{\figurename}{}
\renewcommand{\thefigure}{Fig. S\arabic{figure}}

\setcounter{equation}{0}
\setcounter{figure}{0}

\begin{figure*}
\centering
\includegraphics{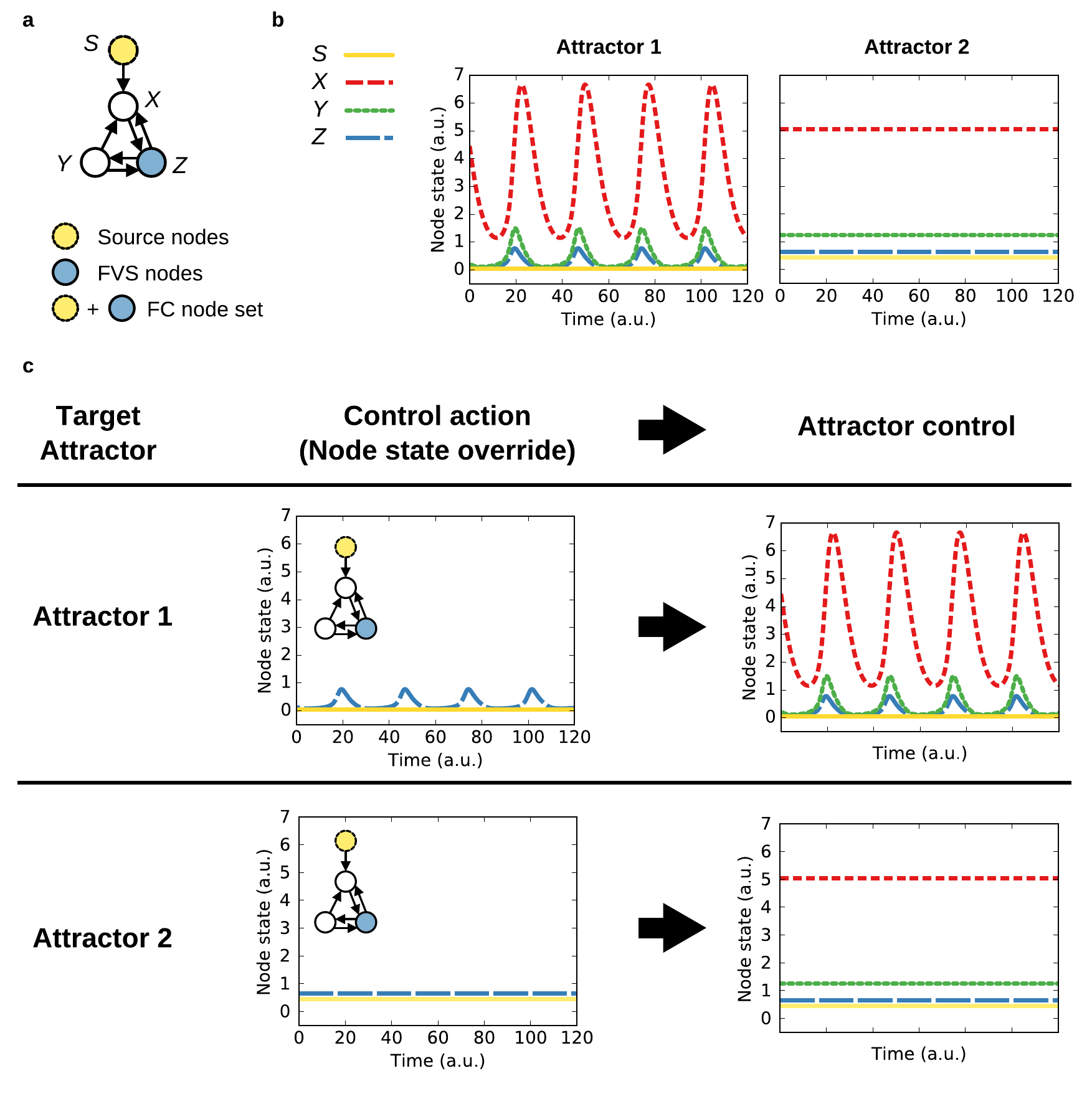}
\caption{Example of Feedback Vertex Set Control. (a) Network representation of the system governed by Eqs. \ref{eq:ex1}-\ref{eq:ex4} and its Feedback Vertex Set Control (FC) node set. (b) Two attractors of the system governed by Eqs. \ref{eq:ex1}-\ref{eq:ex4}, a limit cycle (Attractor 1) and a steady state (Attractor 2). The time course of each node state variable is denoted with different line styles and colors. The colors of the FC node set match those in panel a. (c) For a target attractor of interest (left column), the control action of a node state override of the FC node set (middle column) guarantees that the system will converge to the target attractor (right column).}
\label{fig:FVSParExample}
\end{figure*}

\begin{figure*}
\centering
\includegraphics{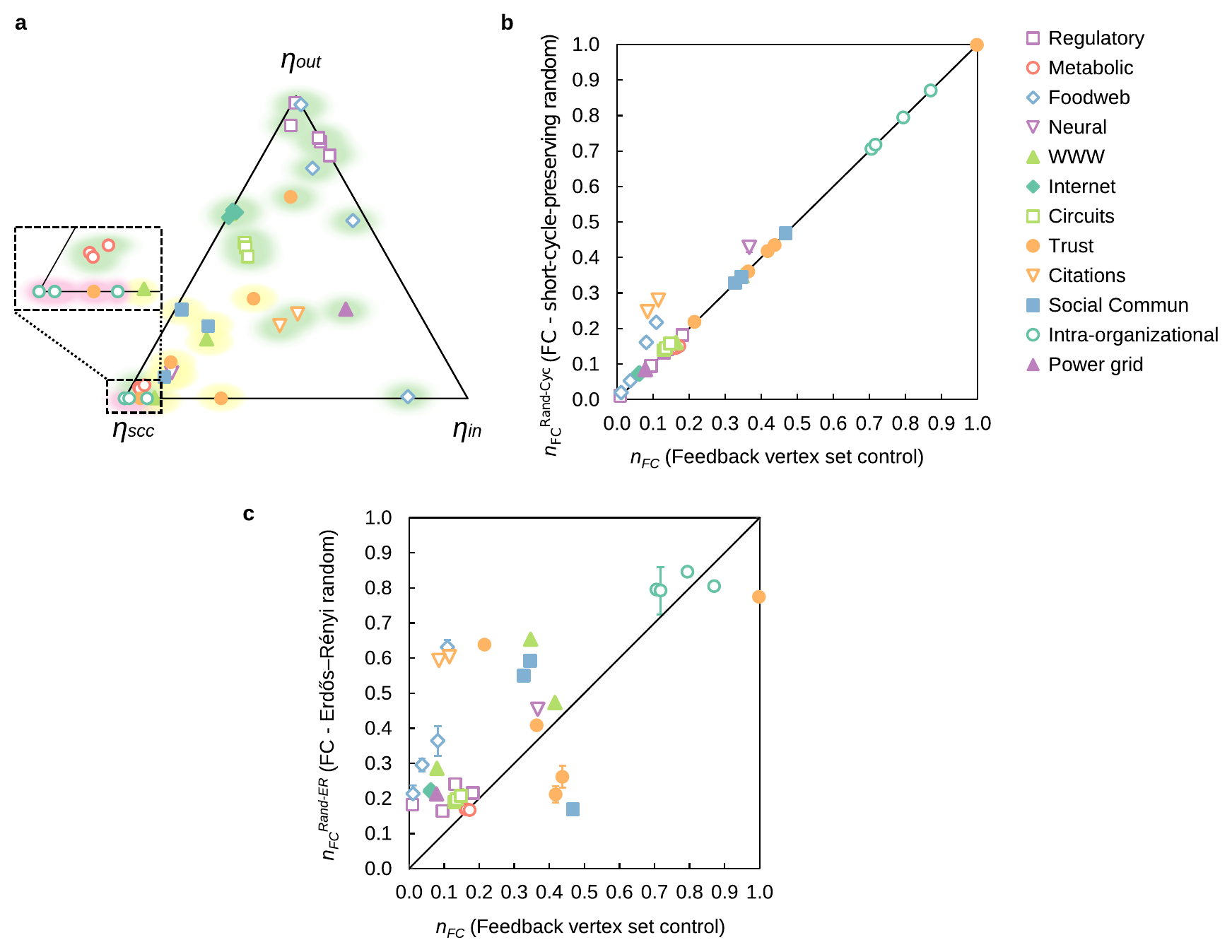}
\caption{Additional results on feedback vertex set control in real networks. (a) Ternary plot of the normalized fraction of nodes in an SCC ($\eta_{SCC}$), in the in-component of all SCCs ($\eta_{in}$), and in the out-component of all SCCs ($\eta_{out}$) for each real network ($\eta_x=N_x/(N_{SCC}+ N_{in}+ N_{out}$), $x=SCC, in, out$). The position in the plot is determined by $\eta_x$ in such a way that a point is close to the $\eta_x$ vertex if $\eta_x \simeq 1$ and close to the side opposite to the $\eta_x$ vertex if $\eta_x \simeq 0$. Thus, networks dominated by their strongly connected component are close to the $\eta_{SCC}$ vertex, networks dominated by their out-component are close to the $\eta_{out}$ vertex, and networks dominated by their in-component are close to the $\eta_{in}$ vertex. The green, yellow and pink shading is defined as in Fig. 2a. Networks with the largest FC node set size (pink shading) are dominated by their SCC, networks with an intermediate FC node set size (yellow shading) have an intermediate-to-high value of $\eta_{SCC}$, while several networks with the lowest FC node set size (green shading) are dominated by their out-component (e.g. regulatory networks) or in-component. Metabolic networks are outliers: they have one of the largest $\eta_{SCC}$ (>0.9) yet have a small $n_{FC}$ (<0.25) and $n_{FVS}$ (<0.2, see Fig. 2a,b ). We attribute this result of metabolic networks to their low connectivity and density ($M/N<3$ and $M/N^2<0.003$, respectively, where $M$ is the number of edges), which makes it easier to disrupt the cycle structure of the large SCC. (b) Scatter plot with the fraction of control nodes in FC for real networks ($n_{FC}$) and their short-cycle preserving randomization ($n_{FC}^{Rand-Cyc}$). (c) Scatter plot with the fraction of control nodes in feedback vertex set (FVS) control for real networks ($n_{FC}$, horizontal axis) and their full randomization (Erd\H{o}s-R\'{e}nyi, $n_{FC}^{Rand-ER}$). $n_{FC}$ in real networks shows a weak correlation with its value $n_{FC}^{Rand-ER}$ in full randomization. The intra-organizational networks at the top-right part of the plot have a large graph density and are close to being complete graphs; because of this, the feedback vertex set of these networks and their Erd\H{o}s-R\'{e}nyi networks is very similar (i.e., the FVS is approximately the whole graph). Error bars denote the estimated standard deviation of the randomized ensembles.}
\label{fig:FVSrealSI}
\end{figure*}

\begin{figure*}
\centering
\includegraphics{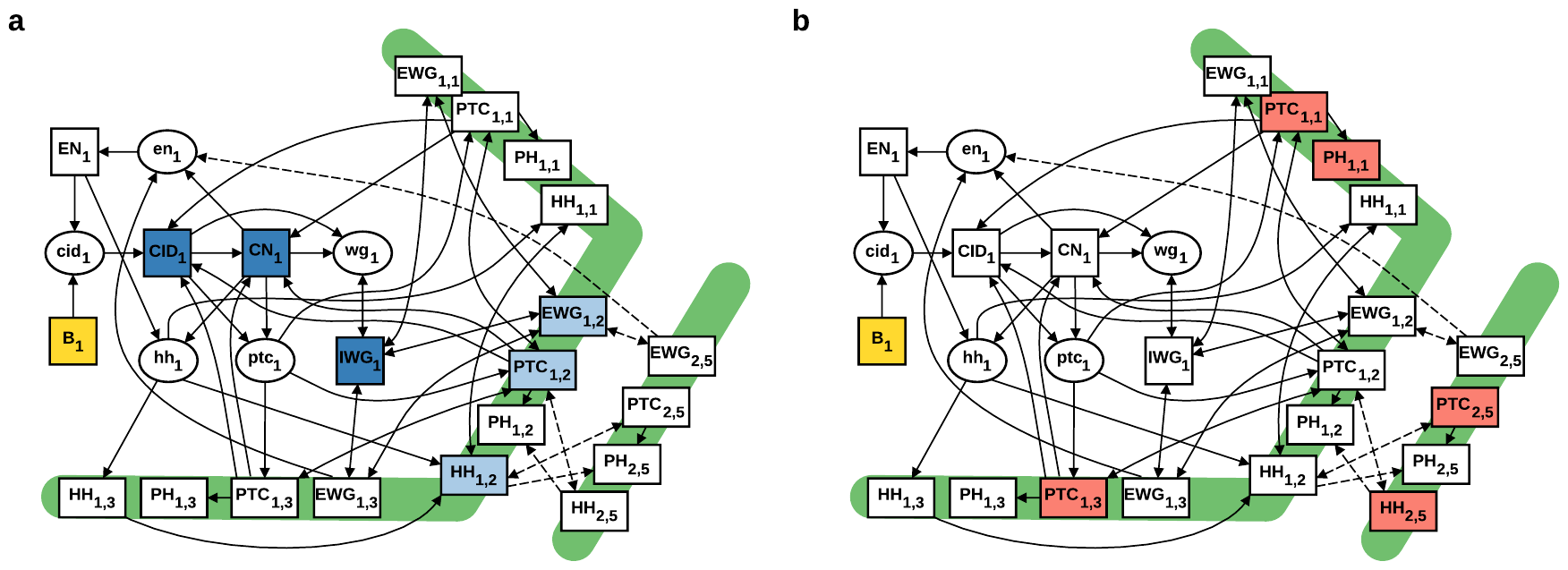}
\caption{Control of the von Dassow et al. model of the Drosophila segment polarity network. The figure shows a cell of the four-cell parasegment together with three of its six boundaries (green lines). The complete network contains four cells in a symmetric completion of the figure. Elliptical nodes represent mRNAs and rectangular nodes are proteins. Intracellular interactions are drawn as solid lines and intercellular interactions are dashed. Yellow nodes are source nodes. (a) Blue nodes are FC nodes in every cell. Dark blue nodes are sufficient for attractor control in the considered dynamic models. (b) Red nodes are SC nodes in every cell.}
\label{fig:SFig2}
\end{figure*}

\begin{figure*}
\centering
\includegraphics{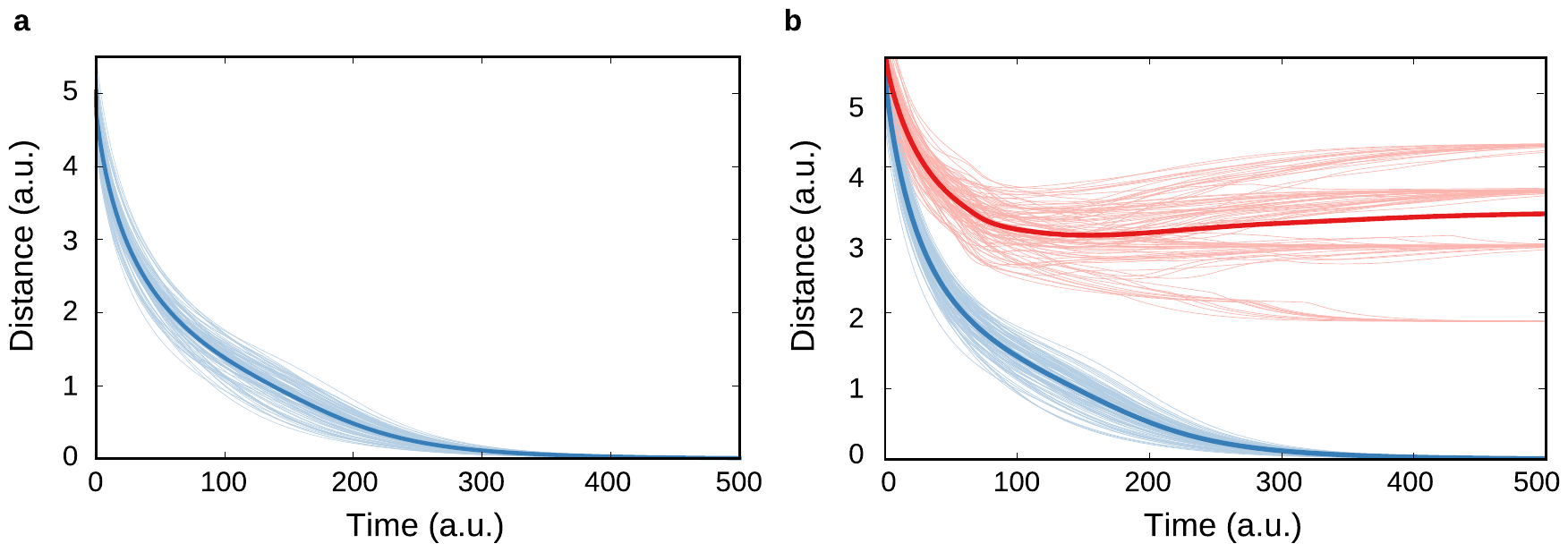}
\caption{Effectiveness of the control of the Drosophila segment polarity differential equation model. (a) The thin light blue lines indicate the evolution of the norm of the difference between the desired wild type steady state and the controlled state trajectory using FC (blue symbols on \ref{fig:SFig2}a) for 100 randomly chosen initial conditions. (b) The thin light blue lines are the evolution of the norm of the difference between the wild type steady state and the controlled state trajectory using reduced FC (dark blue symbols on \ref{fig:SFig2}a) for 100 randomly chosen initial conditions. The thin red lines indicate the norm of the difference between the uncontrolled trajectory and the wild type steady state for 100 randomly chosen initial conditions. In all initial conditions the concentration of each quantity is chosen uniformly from the interval $[0,1]$. The thick blue (red) lines indicate the average of the relevant 100 realizations.}
\label{fig:SFig3}
\end{figure*}

\begin{figure*}
\centering
\includegraphics{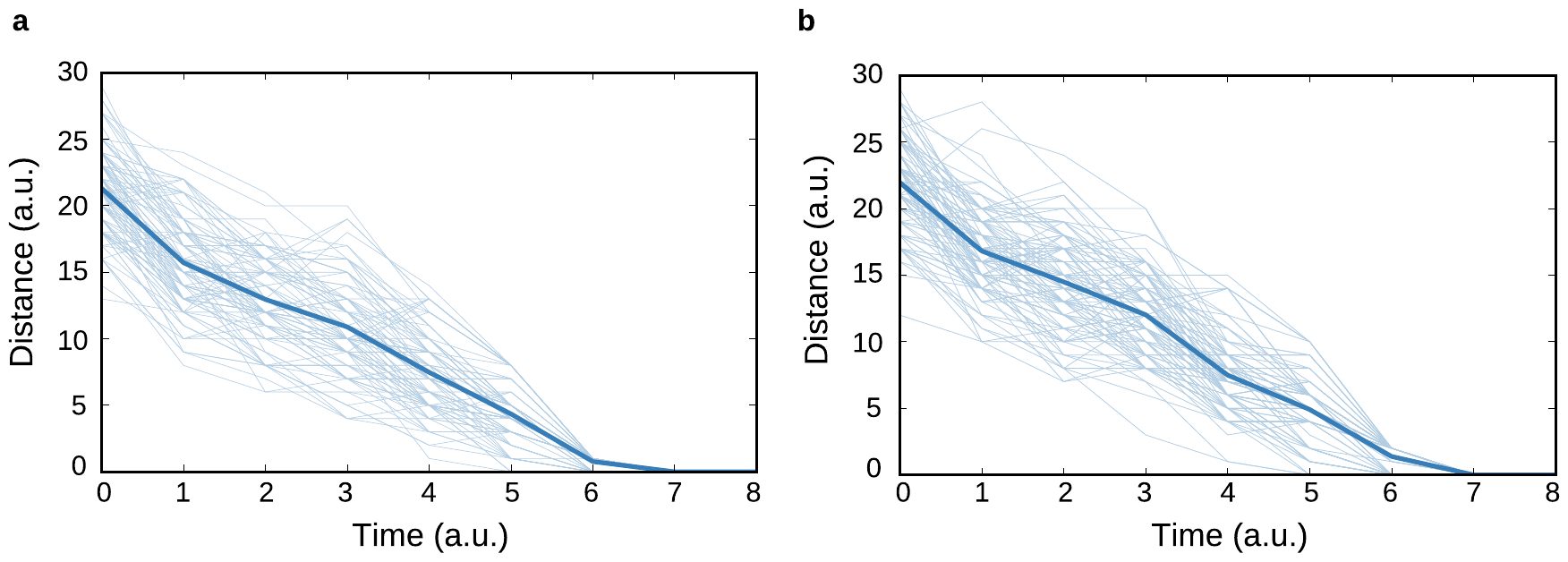}
\caption{Control of the Boolean model of the Drosophila segment polarity genes. The light blue thin lines show the evolution of the norm of the difference between the wild type steady state and the controlled state trajectory using feedback vertex set control (FC) for 100 randomly chosen initial conditions, in which the concentration of each quantity is chosen between ON and OFF with equal odds. The thick blue line is the average of the 100 realizations. (a) Control using the feedback vertex set (b) Control using the reduced feedback vertex set.}
\label{fig:SFig5}
\end{figure*}

\begin{figure*}
\centering
\includegraphics{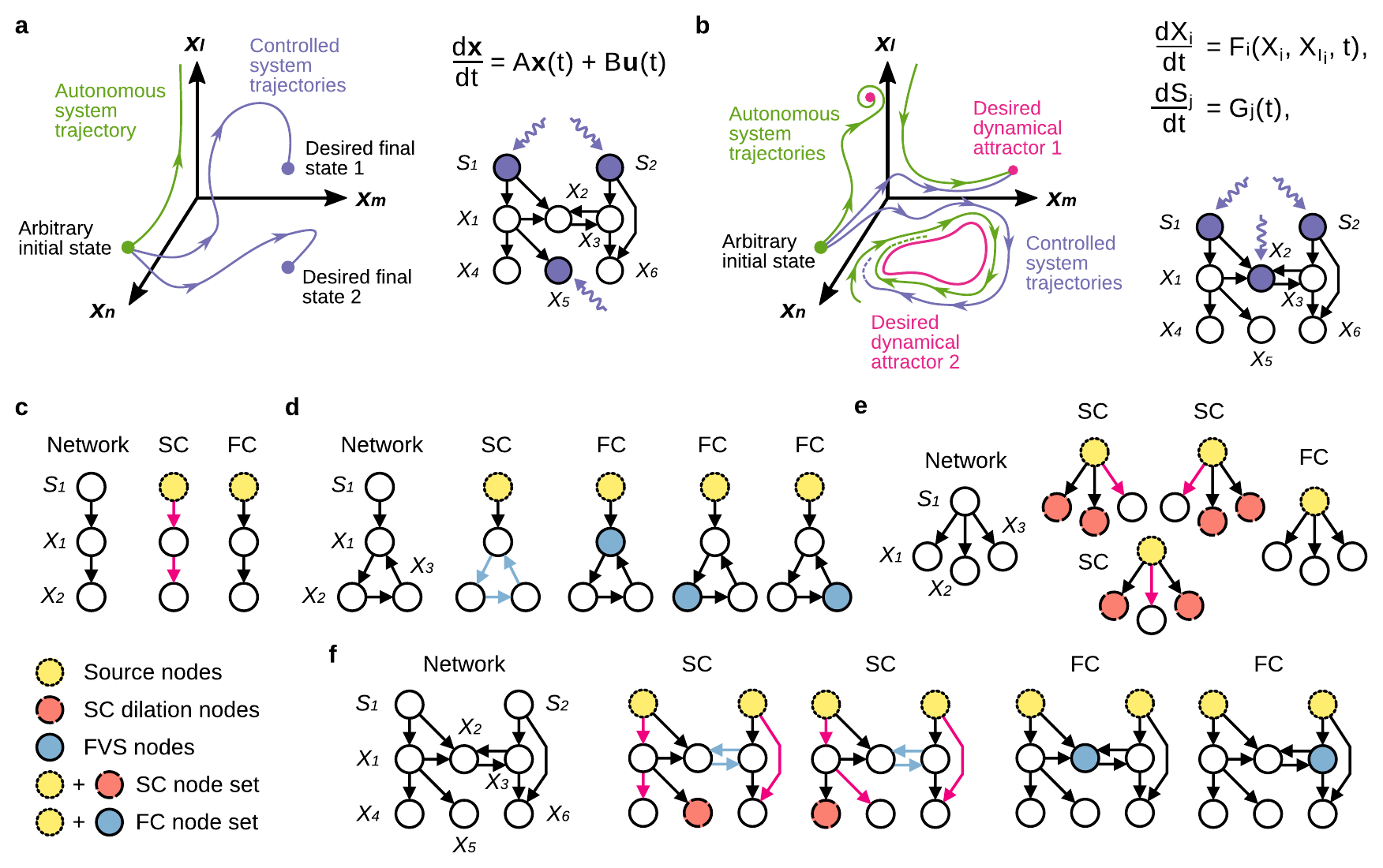}
\caption{Structure-based control methods. Structure-based control methods make conclusions about the dynamics of a system using solely the network structure. This figure repeats some panels from Fig. 1. (a) In structural controllability (SC) the objective is to drive the network from an arbitrary initial state to any desired final state by acting on the network with an external signal $\bf{u}(t)$. The dynamics are considered to be well-approximated by linear dynamics. (b) In feedback vertex set control (FC) the objective is to drive the network from an arbitrary initial state to any desired dynamical attractor (e.g. steady state) by overriding the state of certain nodes. (c-f) Structure-based control in simple networks. Control of the source nodes (yellow nodes with dotted outlines) is shared by SC and FC. SC additionally requires controlling certain dilation nodes (red nodes with dashed outlines) but requires no independent control of cycles. FC requires controlling all cycles by control of the feedback vertex set (FVS, blue nodes with solid outlines). The edges of the non-intersecting linear chains of nodes of SC are colored purple and the edges involved in a directed cycle are colored blue.}
\label{fig:ControlExample2}
\label{fig:ControlFigure2}
\end{figure*}

\begin{figure*}
\centering
\includegraphics{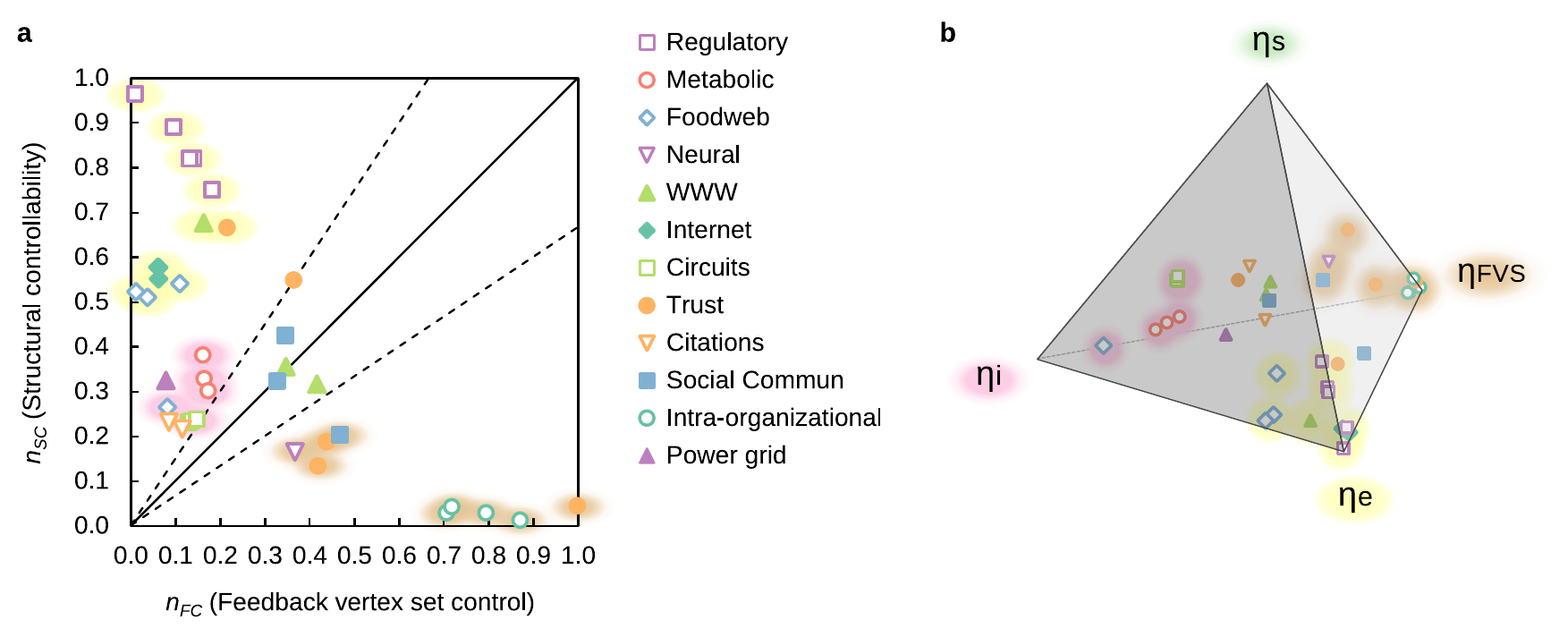}
\caption{Structure-based control in real networks. (a) Scatter plot with the fraction of control nodes in feedback vertex set (FVS) control ($n_{FC}$) and structural controllability ($n_{SC}$) for each real network in \ref{tab:STable2}. The bold line denotes the positions in the plot with $n_{SC}=n_{FC}$, while the dashed lines denote $n_{SC}=1.5 \ n_{FC}$ and $n_{FC}=1.5 \ n_{SC}$. The shading of the symbols corresponds to their position in panel b. (b) Barycentric plot of the normalized fraction of control nodes $\eta_x$, where $x=s, e, i, FVS$ for each real network. The position in the plot is determined by $\eta_x$ in such a way that a point is close to the $\eta_x$ vertex if $\eta_x \simeq 1$ and close to the face opposite to the $\eta_x$ vertex if $\eta_x \simeq 0$. Thus, networks dominated by their FVS and strongly connected component are close to the $\eta_{FVS}$ vertex (brown shading), networks dominated by their out-component are close to the $\eta_{e}$ vertex (yellow shading), and networks dominated by internal dilations are close to the $\eta_{i}$ vertex (pink shading). Networks dominated by their in-component would be close to the $\eta_s$ vertex (green shading), but none of the networks are.}
\label{fig:FVSvsSC2}
\end{figure*}

\begin{figure*}
\centering
\includegraphics{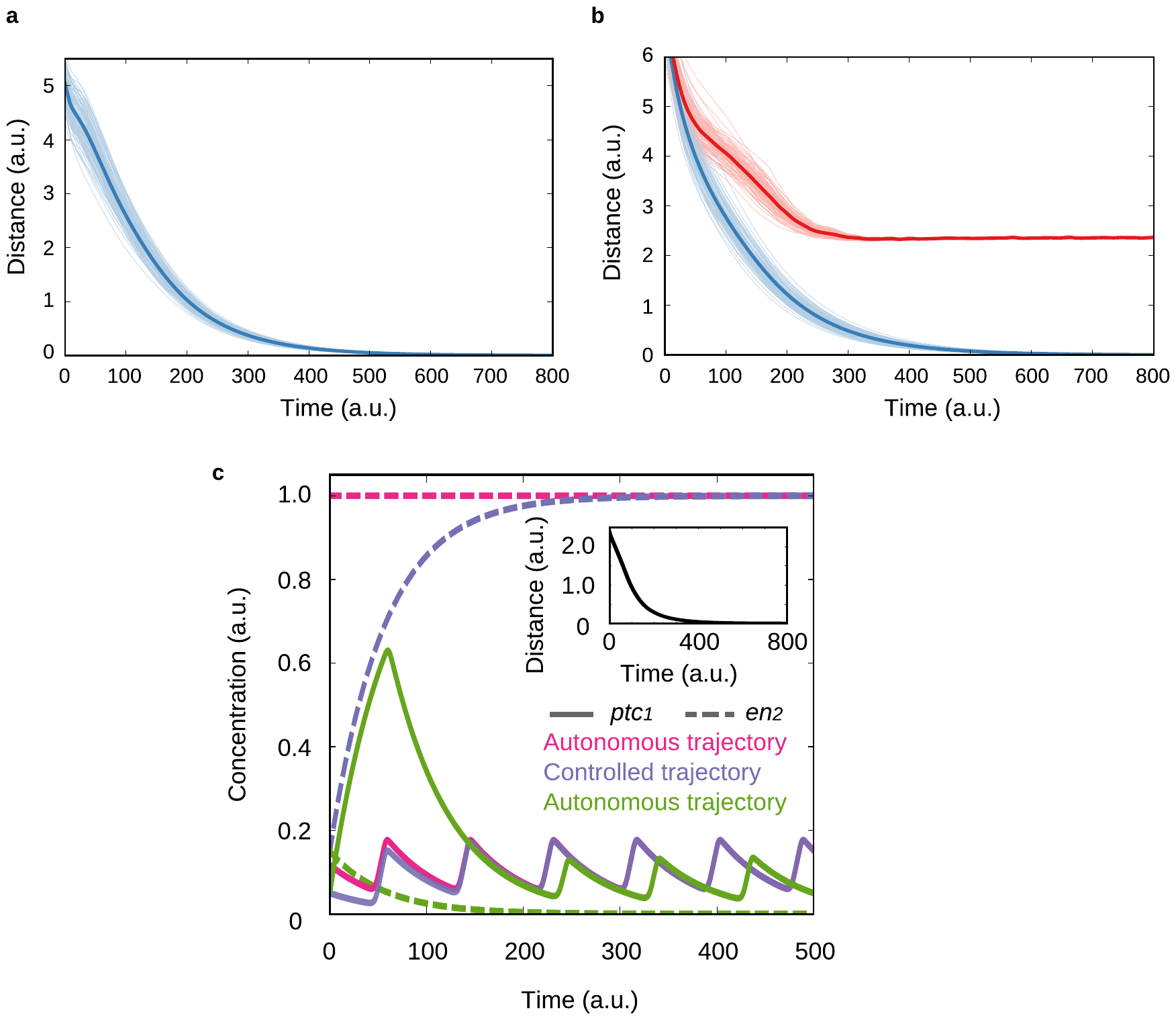}
\caption{Control of the Drosophila segment polarity gene differential equation model for a different parameter set than that used to generate Fig. 4. (a) The thin light blue lines show the evolution of the norm of the difference between the wild type attractor and the controlled state trajectory using FC for 100 randomly chosen initial conditions. (b) The thin light blues lines are the evolution of the norm of the difference between the wild type attractor and the controlled state trajectory using reduced feedback FC for 100 randomly chosen initial conditions. The thin red lines are the evolution of the norm of the difference between the wild type attractor and uncontrolled trajectory using reduced FC for 100 randomly chosen initial conditions. In all initial conditions the concentration of each quantity is chosen uniformly from the interval [0,1]. The thick blue(red) line is the average of the 100 realizations. (c) The concentration of \textit{ptc} in the first cell (solid lines) and en in the second cell (dashed lines) with respect to time. Pink lines and green lines represent autonomous trajectories that start from different initial conditions (a wild type initial condition and a nearly null, respectively) and converge to different attractors (the wild type limit cycle and an unpatterned limit cycle, respectively). Blue lines represent the case when the system starts from the nearly null initial condition, and after applying FC, evolves into the wild type limit cycle. Inset: evolution of the norm of the difference between the desired attractor and the controlled state trajectory using FC.}
\label{fig:SFig4}
\end{figure*}

\begin{table*}
\centering
\caption{Network and control properties of the real networks analyzed. For each network, we show its number of nodes ($N$), number of directed edges ($M$), the fraction of feedback vertex set control (FC) nodes ($n_{FC}$), the fraction of feedback vertex set nodes ($n_{FVS}$), the fraction of source nodes ($n_{s}$), the fraction of nodes in a strongly connected component (SCC) ($n_{SCC}$), the normalized fraction of nodes in a SCC ($\eta_{SCC}$), in the out-component of all SCCs ($\eta_{out}$), and in the in-component of all SCCs ($\eta_{out}$), the sum of the cycle number z-scores, the average fraction of FC nodes in degree-preserving randomized networks ($n_{FC}^{Rand-Deg}$), in SCC-preserving randomized networks ($n_{FC}^{Rand-SCC}$), and in short-cycle-preserving randomized networks ($n_{FC}^{Rand-Cyc}$). The second page of the table shows the number of 1-cycles, 2-cycles, and 3-cycles in real networks, and the mean and standard deviation (S.D.) of the cycle numbers in degree-preserving randomized networks. The z-score of each cycle number is calculated using $(C_L^{Real}-C_L^{Rand})/\sigma_{C_L}$, where $C_L^{Real}$ is the number of L-cycles in the real network, $C_L^{Rand}$ is the mean number of cycles in degree-preserving randomized networks, and $\sigma_{C_L}$ is the standard deviation of the number of cycles. The third page of the table shows the number of 4-cycles,the mean and standard deviation (S.D.) of 4-cycles in degree-preserving randomized networks, the fraction of nodes to be controlled under structural controllability (SC) ($n_{SC}$), the fraction of external nodes ($n_{e}$), the fraction of internal nodes ($n_{i}$), the normalized fraction of feedback vertex set ($\eta_{FVS}$), source ($\eta_{s}$), external ($\eta_{e}$), and internal ($\eta_{i}$) control nodes, and the average fraction of FC nodes in fully randomized (Erd\H{o}s-R\'{e}nyi) networks ($n_{FC}^{Rand-ER}$). (*) The cycle z-score is larger than the number shown; the number of cycles in the real network exceeded $2\times10^6$. (**) The maximum cycle length used was 3 instead of 4 because of the large number of cycles in both the real and randomized networks.}
\includegraphics{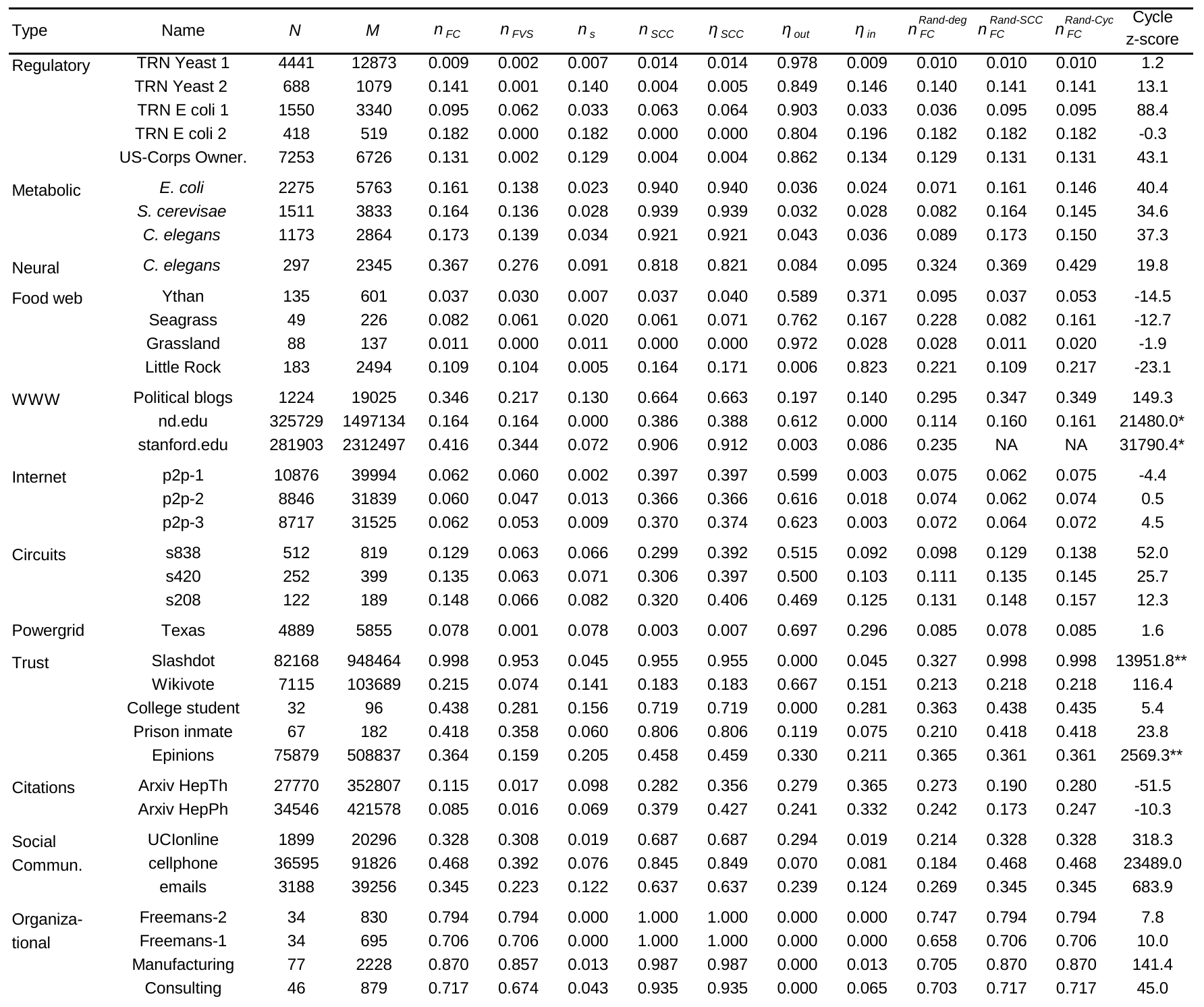}
\label{tab:STable2}
\label{tab:STable}
\end{table*}

\setcounter{table}{0}
\begin{table*}
\centering
\caption{Continuation of \ref{tab:STable2}.}
\includegraphics{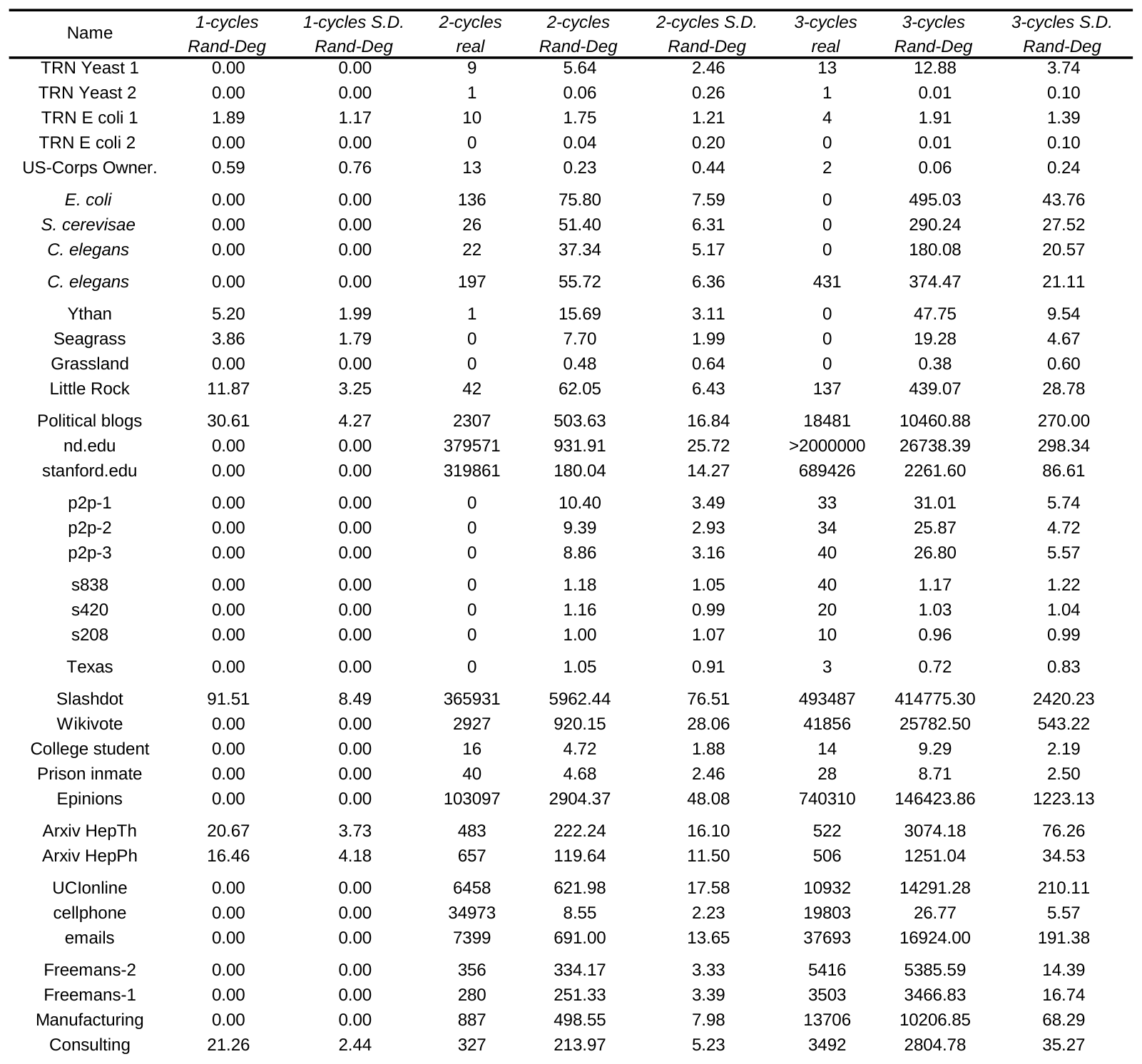}
\end{table*}

\setcounter{table}{0}
\begin{table*}
\centering
\caption{Continuation of \ref{tab:STable2}.}
\includegraphics{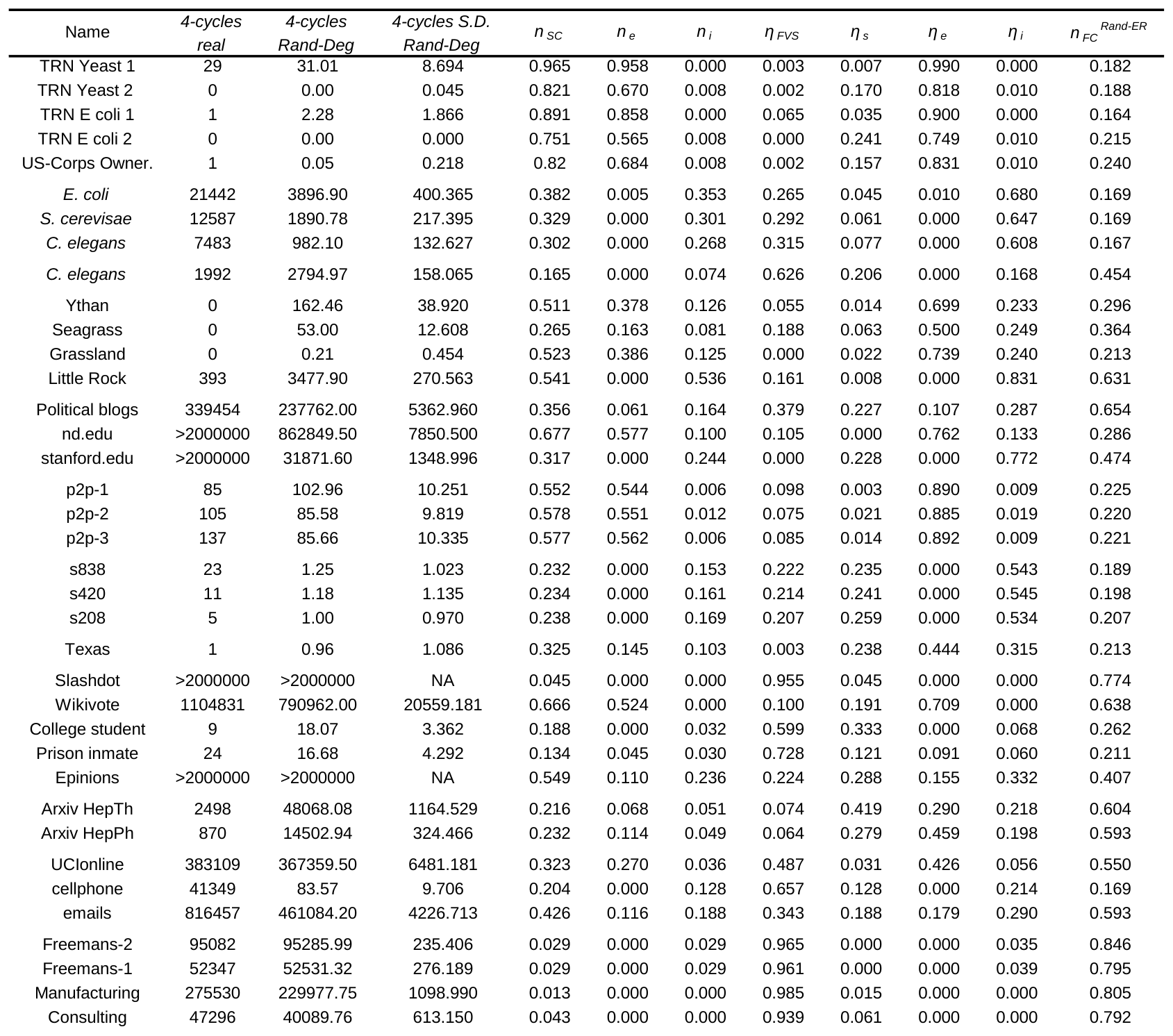}
\end{table*}

\begin{table*}
\centering
\caption{Summary statistics of the distribution of $n_{FVS}$ for real networks using different algorithms. We use two algorithms: the GRASP algorithm \cite{FVSalgorithm1,FVSalgorithm2} (GRASP) and the simulated annealing algorithm of ref. \cite{FVSalgorithm3} (SA). For the GRASP algorithm we use the default parameters, and for the simulated annealing algorithm we use the same parameters as in \cite{FVSalgorithm3} except for a $maxMvt$ value between 0.05-5 times $N$ (the network size) for the inner loop iteration parameter. The number of iterations for GRASP is 2000 for most networks and 50 for some of the largest networks (nd.edu, Slashdot, Epinions, Arxiv HepPh, Arxiv HepTh, and cellphone networks). The number of iterations for SA is between 10 and 100 per network.}
\includegraphics{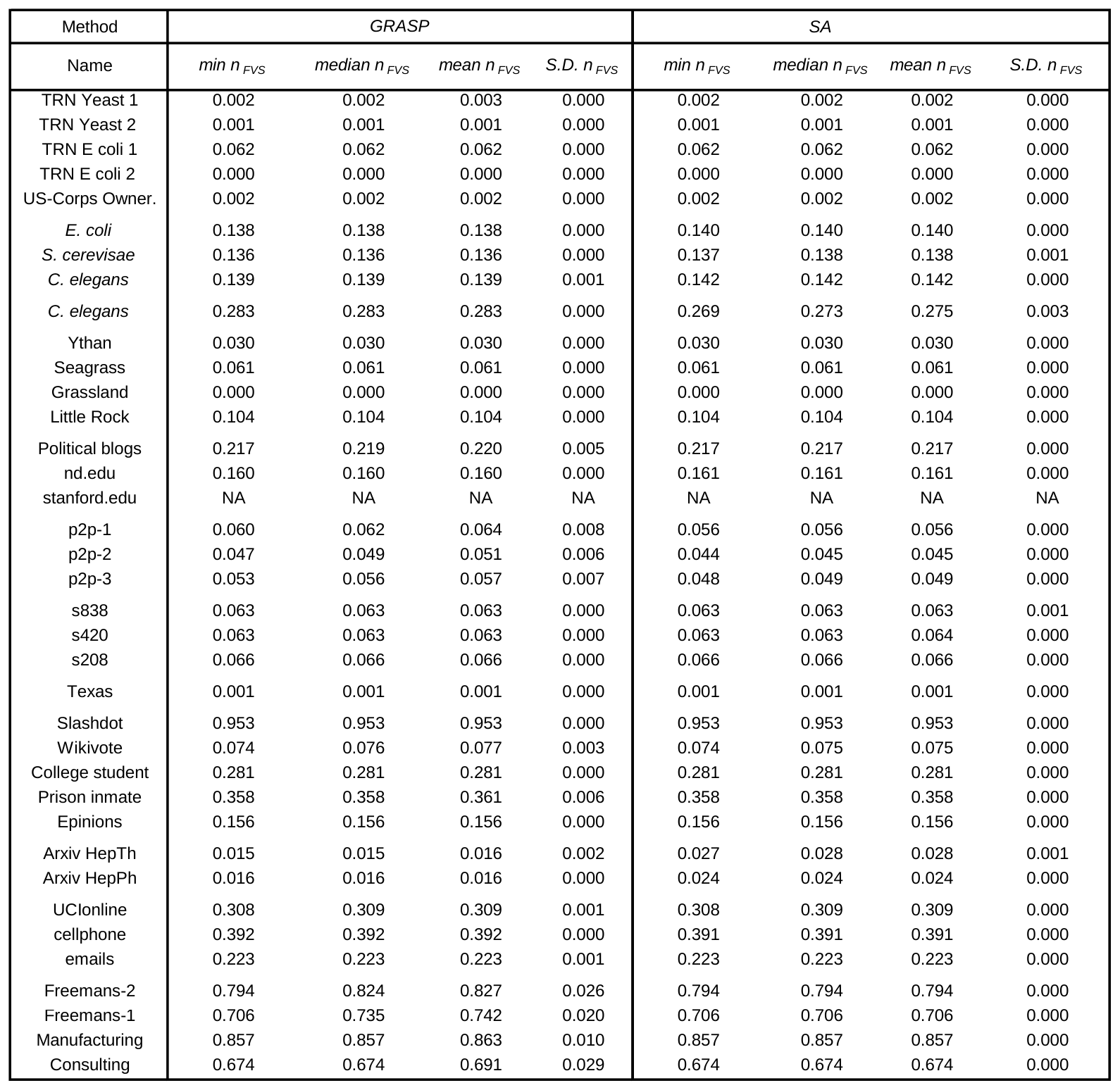}
\label{tab:STable3}
\end{table*}

\begin{table*}
\centering
\caption{Table comparing the control properties of feedback vertex set control (FC) and structural controllability (SC).}
\includegraphics{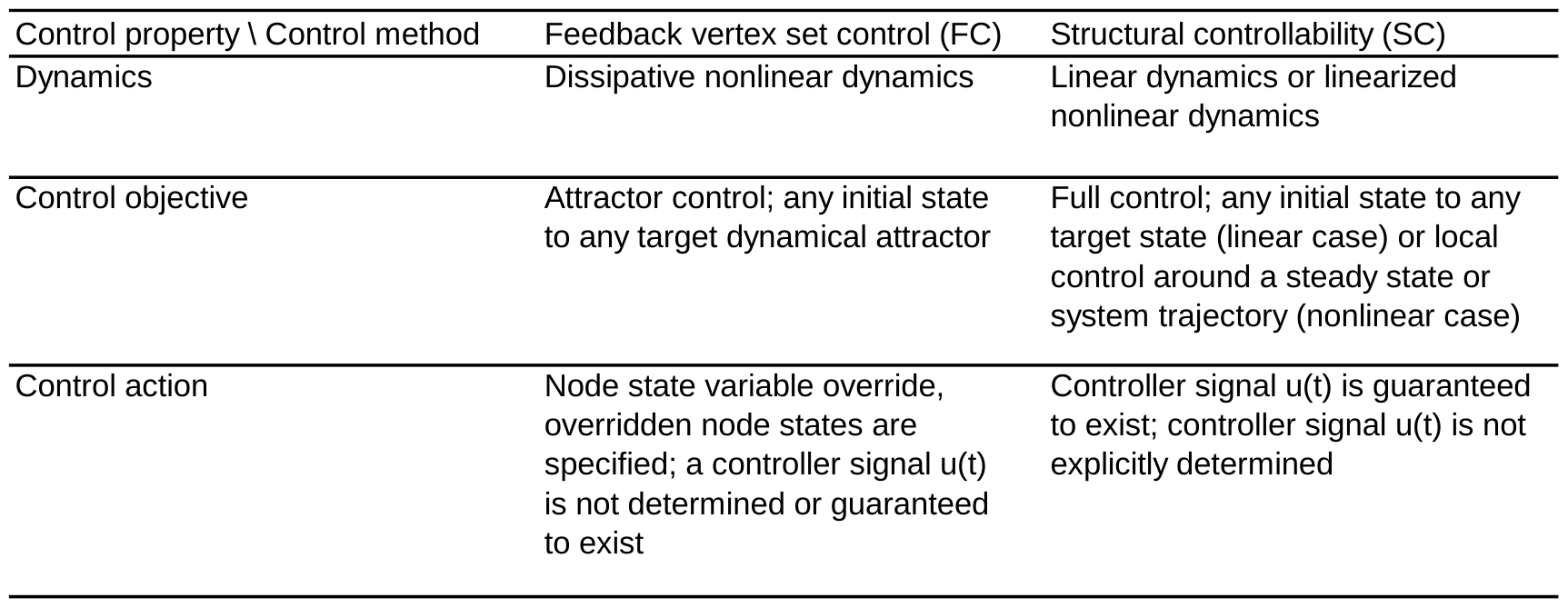}
\label{tab:STable4}
\end{table*}

\clearpage

\section*{SI Text} \label{sec:SText}

\subsection*{I. Feedback vertex set control} \label{sec:PrevFVS}

\subsubsection*{I.A. Previous work on feedback vertex set control} \label{sec:PrevFVS}

In \cite{FVSMath,FVSControl}, Mochizuki, Fiedler et al. introduced the mathematical framework underlying feedback vertex set control (FC). Here we give a brief overview of the main concepts and results of \cite{FVSMath} and its relation the work presented here. In the following $X_i(t)$, $i=1, 2, \ldots, N$, denotes the state of the variable associated to node $i$ at time $t$, and $\textbf{X}=(X_1, X_2, \ldots, X_N)$ is a vector composed of the state of the variables of the network. In addition, we use $X_J$ to denote $X_j$ where $j \in J \subseteq \{1, 2, \ldots, N\}$.

Let each of the system's node states $X_i(t)$ evolve in time according to the differential equations
\begin{align}
\frac{dX_i}{dt} &= F_i(X_i,X_{I_i},t), &i&=1, 2, \ldots, N, \label{eq:s1}
\end{align}
where $F_i(X_i,X_{I_i},t)$ encodes the network structure; $I_i$ defines the predecessor (regulator) nodes of node $i$ in the network and is such that self-loops are included in $I_i$ only if the self-interaction is positive (i.e., $I_i$ contains node $i$ only if $\partial F_i / \partial X_i\geq0$). In other words, negative self-regulation ($\partial F_i / \partial X_i<0$) is not included in $I_i$, only positive self-regulation is \footnote{Note that considering only positive self-regulation as part of $I_i$ is equivalent to adding a new auxiliary variable $\zeta_i$ to encode for positive self-regulation (if any) and not including $i$ as part of $I_i$. In other words, if $\partial F_i / \partial X_i\geq0$ with $i\not \in I_i$, then we introduce $\zeta_i=X_i$ and set $\tilde F_i = F_i(\zeta_i,X_{I_i},t)+\zeta_i-X_i$ as the new equation for node $i$. This would make $\partial \tilde F_i / \partial X_i<0$ for the expanded system and would make the feedback vertex set of the expanded system always include $X_i$ or $\zeta_i$. This approach of adding an auxiliary variable is used in \cite{FVSMath,FVSControl}.}. Furthermore, the $F_i$'s must depend negatively on the first argument of $X_i$ (i.e., they must satisfy the decay condition $\partial_1 F_i (X_i,X_{I_i},t)<0$, where $\partial_1$ indicates the partial derivative with respect to the $X_i$ argument but not the $X_{I_i}$ argument). Additionally, $F_i$ and its first derivatives are assumed to be continuous functions and are assumed to be such that $\textbf{X}(t)$ is bounded ($|\textbf{X}(t)|<C$ for some constant $C$) for any finite initial condition $\textbf{X}(t_0)$ and for all $t\geq t_0$, including the limit $t\rightarrow\infty$. Note that Eq. \ref{eq:s1} determines the dynamics of all node variables, including source nodes, which stands in contrast to Eqs. 1-2 in the main text (Eqs. \ref{eq:sdyn1}-\ref{eq:sdyn2}). We consider the more general case of Eqs. 1-2 in \hyperref[sec:CurrFVS]{Section I.B}.

The boundedness conditions listed in the previous paragraph makes this system a so-called dissipative dynamical system, and guarantee that any initial state will converge to a global attractor $\mathcal{A}$ as $t\rightarrow\infty$,
\begin{align}
\mathcal{A} &= \left\{ \textbf{X}(0) \left| \ \sup_{t\in \mathbb{R}}|\textbf{X}(t)|<\infty \right\} \right..
\end{align}
The global attractor $\mathcal{A}$ is bounded and invariant under Eq. \ref{eq:s1}, and contains all bounded dynamical attractors: steady states, limit cycles, quasi-periodic orbits, and bounded chaotic trajectories.

For the system we consider, the following theorem (Theorem 1.3 in \cite{FVSMath}) forms the basis of FC:

\textit{Theorem. Consider a differential equation system governed by Eq. \ref{eq:s1} with dissipative functions $F_i$, and the associated directed graph $G$ obtained from the $I_i$. We also assume $F_i$ and its derivatives to be continuous. Moreover, $G$ can contain a self-loop only if $F_i$ does not satisfy the decay condition $\partial F_i/ \partial X_i <0$. Then a possibly empty subset $J\subseteq\{1, 2, \ldots, N\}$ of vertices of $G$, and any two solutions $\textbf{X}$ and $\widetilde{\textbf{X}}$ of Eq. \ref{eq:s1} satisfy}
\begin{align*}
\lim_{t\rightarrow\infty} \left(X_J(t) - \widetilde X_J(t)\right) &\rightarrow 0 & \hbox{ implies} \\
\lim_{t\rightarrow\infty} \left(\textbf{X}(t) - \widetilde{\textbf{X}}(t)\right) &\rightarrow \textbf{0}
\end{align*}
\textit{for all choices of nonlinearities $F_i$ if and only if $J$ is a feedback vertex set (FVS) of the graph $G$.}

A consequence of this theorem is that a system governed by Eq. \ref{eq:s1} with an empty FVS must have any pair of solutions approach each other as $t\rightarrow \infty$, i.e., there is single dynamical attractor. Now, if we take a system with a non-empty FVS and override the node state variables of its FVS with their value in the trajectory of any of its dynamical attractors $\mathcal{D}$, then the overridden system is equivalent to a system with an empty FVS \footnote{
Let $J\subseteq\{1, 2, \ldots, N\}$ be the node indices of a FVS, and let $K=\{1, 2, \ldots, N\}/J$ be the node indices of nodes not in the FVS. The dynamics of nodes $K$ in the overridden system are given by $\dot{X}_k=F'_k(X_k,X_{I'_k},t)=F_k(X_k,X_{I_k},t)\mid_{X_J(t)=X_J^{\mathcal{D}}(t)}$, $k \in K$, where $X_J^{\mathcal{D}}(t)$ is the trajectory of the overridden node states. Since $F'_k(X_k,X_{I'_k},t)=F_k(X_k,X_{I_k},t)\mid_{X_J(t)=X_J^{\mathcal{D}}(t)}$, then $I'_k$ does not contain any node in $J$ and the graph defined by the $I'_k$ will have no cycles (removing $J$, by definition, makes the graph acyclic).}. Since the dynamical attractor $\mathcal{D}$ is still a dynamical attractor of the overridden system, which has an empty FVS, it must be the only dynamical attractor of the overridden system. Hence, if we override the dynamics of the FVS of system Eq. \ref{eq:s1} with the trajectory in one of its dynamical attractors, this theorem guarantees that the overridden system will converge to this attractor. Furthermore, overriding the full FVS is necessary and sufficient if one wants this control action to hold for all choices of $F_i$'s.

The FVS framework does not predict what happens when the node state override is close to but not precisely at the prescribed state (likely because this could be model-dependent and attractor-dependent). In general, the expectation is that a node state override that mimics the desired state as closely as possible will move the system into the basin of attraction of the desired attractor and the system will thus converge into the desired attractor.

\subsubsection*{I.B. Feedback vertex set control for general system dynamics} \label{sec:CurrFVS}

Consider the general system used in the main text. The state of the system's $N$ nodes at time $t$, characterized by source node variables $S_j(t)$ (for nodes with no incoming edges) and internal node variables $X_i(t)$, obeys the equations
\begin{align}
\frac{dX_i}{dt} &= F_i(X_i,X_{I_i},t), &i&=1, 2, \ldots, N-N_s, \label{eq:sdyn1} \\
\frac{dS_j}{dt} &= G_j(t), ), &j&=N-N_s+1, \ldots, N, \label{eq:sdyn2}
\end{align}
The dynamics of each source node $j$ is independent of the internal node variables $X_i$ (by definition), and is fully determined by $G_j(t)$, and does not include a decay term. In the simplest case $G_j=0$ and $S_j$ will remain in the specified initial state. The dynamics of each internal node $i$ is governed by $F_i(X_i,X_{I_i},t)$, where the $I_i$ determines the predecessor nodes of $i$ (which can be source or internal nodes) and satisfies the same conditions as in \hyperref[sec:PrevFVS]{Section I.A}. The dynamics are assumed to be bounded, and the $F_i$'s and $G_j$'s and their first derivatives are taken to be continuous.

For this system, the theorem in \hyperref[sec:PrevFVS]{Section I.A} and its consequences (i.e., the results of refs. \cite{FVSControl,FVSMath}) cannot be applied directly since the source node variables $S_j(t)$ do not obey Eq. \ref{eq:s1}. Note that the addition of the source node variables $S_j$ is not merely cosmetic; the $S_j$'s can denote external stimuli the system is subject to or initial-condition-specified node variables (as would happen if $G_j=0$); these stimuli or initial/boundary variables can affect the dynamical attractors available to the system (e.g. steady states can merge or disappear if $S_j$ takes different values, see e.g. \cite{TysonReview,SocialBassett}).

Here we adapt the previous results of feedback vertex set control to the more general system dynamics. Let $\mathcal{D}$ be the desired dynamical attractor and let $S^{\mathcal{D}}_j(t)$ be the source node trajectory in which this attractor is obtained. Now, assume that the system's source nodes are driven by an arbitrary $G_j(t)$. If starting at time $t_0$, we override the state of the source nodes $S_j(t)$ with $S^{\mathcal{D}}_j(t)$, then for $t>t_0$ we will have $S_j(t)$ be in their state in $\mathcal{D}$. Additionally, the dynamics of the $X_i$ for $t>t_0$ can be described by $\dot{X}_i=F_i(X_i,X_{I'_i},t)=F_i\mid_{S_j(t)=E^{\mathcal{D}}_j(t)}$, where the $F_i'$ no longer depend on $S_j$ (i.e., $I'_i$ is $I_i$ with all the $S_j$ removed). Since the dynamics of the modified system now obey Eq. \ref{eq:s1} (with $F'_i$ instead of $F_i$), then we can guarantee that the $FVS$ can be used to steer the system to any dynamical attractor of interest. Finally, since $F'_i=F_i\mid_{S_j(t)=S^{\mathcal{D}}_j(t)}$, then $\mathcal{D}$ is one of the attractors of the modified system ($\dot{X_i}=F'_i$ and $\dot{X_i}=F_i$ with $S_j(t)=S^{\mathcal{D}}_j(t)$ both have the same governing equations). The result is that the overriding the state of the source nodes $S_j$ and of the $FVS$ into the state in a dynamical attractor $\mathcal{D}$ is guaranteed to steer the system to $\mathcal{D}$ as $t\rightarrow \infty$.

As an example, consider the network in \ref{fig:FVSParExample}a, and the governing equations: \begin{align}
\frac{dS}{dt} &= G_S, \label{eq:ex1} \\
\frac{dX}{dt} &= k_x (Z-\alpha_x X), \label{eq:ex2}\\
\frac{dY}{dt} &= S + \frac{X+Z}{1+k_S S}-\alpha_Y Y, \label{eq:ex3}\\
\frac{dZ}{dt} &= \beta_Z+\frac{X^2}{X^2+1} - \alpha_z ZY, \label{eq:ex4}
\end{align}
where $G_S=0$, $k_x=10$, $\alpha_x=0.5$, $k_S=5$, $\alpha_Y=0.2$, $\beta_Z=0.01$, and $\alpha_z=0.2$. Under these conditions, the system has several attractors, including a limit cycle (\ref{fig:FVSParExample}b, Attractor 1) and a steady state (\ref{fig:FVSParExample}b, Attractor 2). FC guarantees that for either of these two attractors, and any others that exist, the control action of overriding the state variables of the FC node set into the trajectory of a target attractor guarantees that any initial state will converge to said attractor. This means that forcing $S$ and $Z$ into the trajectory specified by Attractor 1 guarantees that the rest of the system ($X$ and $Y$) will converge to Attractor 1, and the same is true for any target attractor \ref{fig:FVSParExample}c. Furthermore, if one modifies the parameters or the functional form of the equations of the system, forcing $S$ and $Z$ into the trajectory of an attractor of the modified system guarantees that the modified system will converge to the modified attractor (as long as the modified system is still a dissipative nonlinear system).

\subsubsection*{I.C. Identifying the minimal feedback vertex set control set of a network} \label{sec:IdFVS}

The FC node set of a network of $N$ nodes is composed of the source nodes of the network ($N_s$ of them) and of the FVS of the network. The minimal FC node set $N_{FC}$ of a network is obtained by finding a minimal FVS, since the number of source nodes $N_s$ is fixed for any given network. The minimal FVS of a network is not guaranteed to be unique, and is often found to have a large degeneracy (see the examples in Fig. 1 of the main text).

In order to find the minimal FVS control set of a network, we must find which of the possible $2^{N-N_s}$ node sets is a minimal FVS. The problem of identifying the minimal FVS has a long history in the area of circuit design \cite{FVSReview}. Even though solving the minimal FVS problem exactly is NP-hard \cite{NPhard}, a variety of fast algorithms exist to find close-to-minimal solutions \cite{FVSReview,FVSapprox}. Here we use the FVS adaptation of a heuristic algorithm known as the greedy randomized adaptive search procedure (GRASP) \cite{GRASP}, which is commonly used for combinatorial optimization problems \cite{FVSReview}.
GRASP is an iterative procedure in which each iteration consists of two phases: a construction phase in which a feasible solution to the problem is produced based on a greedy measure and a randomized selection process (given a cutoff for the greedy measure, a feasible solution below the cutoff is chosen randomly and uniformly), and a local search phase in which the local neighborhood in the space of solutions is explored to find a local minimum of the problem. The FVS adaptation of GRASP incorporates the wiring diagram of the network into the procedure by using the in-degree and out-degree of each node as the greedy measure in the construction phase and by utilizing a graph reduction technique that preserves the FVS during the local search phase \cite{FVSalgorithm1,FVSalgorithm2}. In addition, we preprocess all networks by iteratively removing source and sink nodes (this is done iteratively because new source/sink nodes may appear after a source/sink node is removed), since a minimal $FVS$ of a network is invariant under removing nodes that do not participate in directed cycles.

For this work, we use a custom code in Python to iteratively remove source and sink nodes in each network analyzed. The resulting network is then used as an input to the FORTRAN implementation of the FVS adaptation of GRASP \cite{FVSalgorithm1,FVSalgorithm2} using the default settings (2048 iterations and a random uniformly chosen cutoff for the randomized selection process in each iteration), unless otherwise noted.

The NP-hardness of the minimal FVS problem is a limitation of FC, given the approximate nature of any algorithm that can be used on large networks. To evaluate our confidence in the minimal FVS we obtained with the GRASP algorithm\cite{FVSalgorithm1,FVSalgorithm2}, we characterized the distribution of outcomes in all networks (except for the stanford.edu network due to time and resource limitations). The almost identical $n_{FVS}$ obtained using either the minimal or the median result of all iterations (\ref{tab:STable3}) indicates that increasing the number of iterations of the GRASP algorithm would have a small effect on our results. In addition, we also use another algorithm to solve the minimal FVS problem, a simulated annealing algorithm with a novel local search procedure \cite{FVSalgorithm3}. The resulting $n_{FVS}$'s are almost identical (\ref{tab:STable3}), which indicates that our results are not method-dependent. The consistency between these results greatly increases the confidence in the $n_{FVS}$'s we obtained with the GRASP algorithm.

\subsubsection*{I.D. Feedback vertex set control and model-based network control}

Feedback vertex set control gives a set of nodes whose control is sufficient for attractor control in the ensemble of all models that have a given network structure. This set of nodes is also necessary if one demands that control of the same node set be sufficient in every model of the ensemble. Thus, FC gives a sufficiency prediction about the entire ensemble of models with a given network structure, and any particular model in the ensemble may require a smaller set of nodes for attractor control (i.e. the FC node set gives an upper bound for any particular model). Consequently, a subset of the FC node set of a network can often be sufficient for a particular instance of a model with this underlying network structure (section ``Feedback vertex set control and dynamic models of real systems'' of the main text). The generality of this result is supported by a recently developed network control method for Boolean dynamic models called stable motif control \cite{MotifControl}.

Stable motif control is an attractor-based control method that is based on identifying subnetworks which uniquely determine an attractor of interest. Specifically, stable motif control identifies a state manipulation of certain nodes in the subnetworks (namely, fixing them in their states in the desired attractor) that drives any initial state to the desired attractor with 100\% effectiveness \cite{MotifControl}. Stable motif control and feedback vertex set control differ in the dynamic variables they consider (Boolean vs continuous\footnote{Boolean dynamics are a type of nonlinear and dissipative dynamics, which assume two discrete node states, and which can be considered a limiting case of sigmoidal regulatory functions often observed in biological systems, for example, Hill functions with a large Hill coefficient. Boolean dynamics require feedback loops for multi-stability or oscillatory behavior, which guarantees that an acyclic Boolean network (which is equivalent to what we obtain when overriding the state of the FVS) has a unique attractor \cite{Remy}. This is sufficient to guarantee that FC implies attractor control if the FC nodes are fixed (i.e. do not oscillate) in the attractor of interest, but it is not clear if this also extends to oscillating nodes.}) and the information they require (model-based vs structure-based), but they share their attractor-based control objective and their ability to drive any initial state to a desired attractor. Additionally, they share specific methodological aspects:

\begin{description}
\item[(i)] In stable motif control, source nodes are assumed to be fixed in the node state specified by the attractor of interest; if this were not the case, the source nodes would first need to be fixed into the appropriate node state. In FC, source nodes must also be locked in the trajectory specified by the attractor.

\item[(ii)] In stable motif control, each subnetwork identified is either a self-sustaining positive feedback loop (directed cycle), or an intersection of several self-sustaining positive feedback loops. In FC, every feedback loop in the network (both positive and negative) must be manipulated, something which is achieved using an override of the states of the feedback vertex set, which by definition contains a node in every feedback loop in the network.
\end{description}

The first point shows that the treatment of source nodes is almost identical in both methods, and that FC is more general because it allows source nodes to be in any dynamical trajectory and not only a fixed node state, like in stable motif control. The second point shows that the state manipulation of cycles underlies both methods, and that FC requires manipulating all cycles while stable motif control only requires manipulation of a select few positive cycles. The similarities in points (i) and (ii) strongly suggest that stable motif control is the model-based equivalent of feedback vertex set control for the case of Boolean dynamics. In particular, point (ii) gives an explanation of why a reduced FVS can often be sufficient for FC; even though all cycles must be controlled in structure-based control, in a particular model instance only a subset of the cycles (and thus, a subset of the FVS) is sufficient for attractor-based control.

\subsubsection*{I.E. Feedback Vertex Set control and self-dynamics}

In FC, the graph structure (encoded in $I_i$) only needs to consider positive self-loops. This means that FC benefits from knowing some information about the sign (regulatory effect) of self-interactions. Edge sign is otherwise not encoded in the graph structure. If the sign of a self-interaction is not known (or if it can change sign depending on the value of other regulators, as in the case of a logical XOR function), then the self-interaction needs to be included in the graph.

For the case of biological systems, one often knows the regulatory effect of self-interactions from biological evidence (e.g. the regulonDB transcriptional regulatory network in http://regulondb.ccg.unam.mx/ contains the positive, negative or dual nature of interactions), and one can include this in the graph to increase the predictive power of FC. For other biological, social, or technological networks it might not be obvious whether self-interactions are positive, so one needs to choose whether such self-interactions can be positive (which would mean including such self-interactions in the graph) or can only be negative (in which case they do not need to be included in the graph).

\subsubsection*{I.F. Feedback vertex set control and controllers}

In feedback vertex set control we use the control action of forcing (overriding) the state variables into a certain trajectory, namely, the one specified by an attractor of the system we are interested in. The control action of state variable override in FC stands in contrast with the control actions often considered in control theory, in which a controller or driver signal $u(t)$ is coupled to the governing equations $F_i$ and $G_j$, and through this coupling are the trajectories of the state variables $X_i$ and $S_j$ modified. In the simplest case, known as control-affine systems, we would have $\dot{X}_i= F_i + u_i(t) g_i(X,S)$ as the governing equation of the variable of the nodes $i$ we chose to control (and similarly for the source node variables $\dot{S}_j$).

The problem of designing a controller $u(t)$ for nonlinear systems has been the subject of much research over the last decades (e.g. \cite{Slotine,LiuReview,Isidori}), yet designing a controller for a general nonlinear system that drives an initial condition to a target attractor of the system (attractor-based control) is a difficult and unsolved problem (see e.g. section V of the review in \cite{LiuReview}). Recent efforts in attractor-based control require a parameterized model in order to be applicable (and, thus, are not structure-based methodologies) and rely on numerical simulations to design the controller \cite{LiuReview}. For example, refs. \cite{MotterControl} and \cite{GeomControl} give algorithms to numerically obtain a controller (an infinitesimal change of the given initial condition in \cite{MotterControl}, or a temporary modification of parameters in \cite{GeomControl}) that drives the system to the basin of attraction of a target attractor, and leads the system to this attractor

These examples of attractor-based control illustrate that designing a controller for a nonlinear system is a research endeavor of its own and seems to depend strongly on the dynamic model and its parameters. Given that our work focuses on the structure-based aspect of the attractor-control problem, we consider designing a controller to be outside the focus of our current work and a topic of future research. Having said this, the control action of node override can be viewed as an idealized controller signal that allows us to identify the nodes that need to be controlled in systems for which we do not have a parameterized dynamic model or know the system-specific coupling of the controller signal with the equations of motion (Eqs. \ref{eq:sdyn1}-\ref{eq:sdyn2}). Thus, override control is a necessary first step in the task of designing driver signals for these systems.

\subsection*{II. Structural controllability}

\subsubsection*{II.A. Notes on structural controllability}

In structural controllability (SC) we consider a system with an underlying network structure whose autonomous dynamics are governed by linear time-invariant ordinary differential equations
\begin{align}
\frac{d\textbf{x}}{dt} &= A\textbf{x}(t), \label{eq:linaut}
\end{align}
where $\textbf{x}(t)=(x_1(t), x_2(t), \ldots, x_N(t))$ denotes the state of the system, and $A$ is a $N\times N$ matrix that encodes the network structure and is such that $a_{ik}$ is nonzero only if there is a directed edge from $k$ to $i$. Given this system, SC's aim is to identify external driver node signals $\textbf{u}(t)=(u_1(t), \ldots, u_M(t))$ that can steer the system from any initial state to any final state in finite time (i.e., full control, \ref{fig:ControlFigure2}a), and that are coupled to Eq. \ref{eq:linaut} in the following way
\begin{align}
\frac{d\textbf{x}}{dt} &= A\textbf{x}(t) + B\textbf{u}(t), \label{eq:lincont}
\end{align}
where $B$ is a $N\times M$ matrix that describes which nodes are driven by the external signals $\textbf{u}(t)$.

The work of Lin, Shields, Pearson, and others showed that if such a system can be controlled in the specified way by a given pair $(A, B)$, which can be verified using Kalman's controllability rank condition \footnote{Namely, that the $N\times NM$ matrix $(B, AB, A^2B, \ldots, A^{N-1}B)$ has full rank, i.e., $\text{rank}(C)=N$ \cite{Kalman}.}, this will also be true for almost all pairs $(A, B)$ (except for a set of measure zero) \cite{Lin,Shields,Slotine}. In other words, SC is necessary and sufficient for control of almost all linear time-invariant systems consistent with the network structure in $A$. The applicability of SC also extends to nonlinear systems; SC of the linearized nonlinear system around a steady state or system trajectory of interest is a sufficient condition for local controllability of the system around said steady state or trajectory in a sufficiently small time \cite{LiuReview,Slotine,Isidori}. Furthermore, SC of the linearized nonlinear system is also a sufficient condition for some nonlinear notions of controllability such as accessibility \cite{LiuReview,Slotine,Isidori}.

SC is a mathematical formalization of the idea that a node can fully manipulate only one of its successor elements at a time and that a directed cycle is inherently self-regulatory. A consequence of this is that the driver nodes are such that every network node is either part of a set of non-intersecting linear chains of nodes that begin at the driver nodes or is part of a set of directed cycles that do not intersect each other or the set of linear chains and which are reachable from the driver nodes (\ref{fig:ControlFigure2}). As Ruths \& Ruths showed \cite{RuthsControl}, this implies that there are three types of network nodes that must be directly manipulated by a unique driver node, and which we call SC nodes: (i) every source node, and every successor node of a dilation (when a node has more than one successor node) that is not part of the set of linear chains or of the cycles, namely (ii) the surplus of sink nodes with respect to source nodes or (iii) internal dilation nodes.

To illustrate how the nodes that need to be manipulated in SC and FC can differ from each other, consider the example networks in \ref{fig:ControlExample2}. In a linear chain of nodes (\ref{fig:ControlExample2}c, left) the only node that needs to be controlled in both frameworks is the source node $S_1$. For \ref{fig:ControlExample2}d, which consists of a source node connected to a cycle, SC requires controlling only the source node $S_1$ since the cycle is considered self-regulating (\ref{fig:ControlExample2}d, middle), while FC additionally requires controlling any node $X_i$ in the cycle, the feedback vertex set in this network (\ref{fig:ControlExample2}d, right). \ref{fig:ControlExample2}e consists of a source node with three successor nodes; SC requires controlling two of the three successor nodes because of the dilation at the source node $S_1$, while for FC controlling $S_1$ is sufficient. In \ref{fig:ControlExample2}f we show a more complicated network with a cycle and several source and sink nodes, and two minimal node sets for SC and FC. These examples illustrate that the control of the source nodes is shared by full control in SC and attractor control in FC, and that their main difference is in the treatment of cycles, which require to be controlled in FC and do not require independent control in SC.

\subsubsection*{II.B. Structural controllability and self-dynamics}

In a system governed by Eq. \ref{eq:lincont}, self-dynamics is captured by having the matrix elements in the diagonal of $A$ be nonzero (i.e., a self-loop in the network structure). If each node variable in the system has self-dynamics, then every node in the associated graph structure of $A$ will have a self-loop. Directly applying SC to such a graph will yield the surprising result that a single driver signal $\textbf{u}(t)=(u_1(t))$ is necessary and sufficient for full control, regardless of any other aspect of the graph structure \cite{LiuControl,LiuReview,Bergstrom}. This result, although mathematically correct, gives little insight into the impact of the underlying network structure of $A$ (other than self-loops) on control-related questions. Furthermore, as Sun et al. showed using minimal-energy control driver signals \footnote{Minimal-energy control driver signals are the ones that minimize the functional $\int_0^{t_f} ||\textbf{u}(t)||^2 dt$, where $t_f$ is the desired final time.}, the required driver signal $\textbf{u}(t)=(u_1(t))$ might be numerically impossible to implement unless the number of control nodes is significantly increased \cite{SunControl}.

We should emphasize that controllability of a system with self-dynamics by a single driver signal is a consequence of SC's assumption that each nonzero entry in $A$ and $B$ is independent of each other. Thus, if one considers SC for the set of $(A, B)$'s in which the diagonal elements of $A$ are fixed (i.e., the self-dynamics are fixed but every other nonzero entry is still arbitrary) then the number of driver nodes can be obtained from the eigenvalues of $A$ and their geometric multiplicities \footnote{The geometric multiplicity $\mu(\lambda)$ of an eigenvalue $\lambda$ of $A$ is given by $\mu(\lambda)=N-\text{rank}(\lambda I - A)$, where $I$ is the $N \times N$ identity matrix.)}, as shown in a recent study by Zhao et al. \cite{Selfdyn}. For most cases, obtaining the eigenvalues of $A$ and their geometric multiplicities is computationally demanding and requires specifying a value for the weight $a_{ii}$ of each self-loop. For the special case of a single fixed weight $\alpha$ for the self-dynamics of every node ($a_{ii}=\alpha$, $\forall i$), the number of driver nodes is equivalent to the one specified by SC using $A$ but setting all diagonal elements to zero \cite{Selfdyn}.

These considerations about self-dynamics are crucial when using SC on the nonlinear systems we consider, Eqs. \ref{eq:sdyn1}-\ref{eq:sdyn2}. Since the nonlinear functions $F_i$ have a decay term that prevents the system from increasing without bounds, then a linearization of the $F_i$'s will give nonzero diagonal entries for $A$. Thus, SC would predict that a single driver signal is sufficient for controllability regardless of the topology of the real network considered, a result which tells us little about structure-based control in these networks. Instead, we follow the approach of Liu et al. \cite{LiuControl} and do not include the decay self-dynamics as a self-loop in the graph structure. Two equivalent interpretations of this approach under SC are that (i) we consider the decay terms to not dominate the linearized dynamics (i.e., we set them to zero), or (ii) every element has the same (or very similar) fixed weight for its self-dynamics (i.e., the self-dynamics are fixed and every other nonzero entry in $A$ is arbitrary).

\subsubsection*{II.C. Identifying the minimum number of driver nodes in structural controllability} \label{sec:IdFVS}

Here we use the maximum matching approach of Liu et al. \cite{LiuControl} to identify the minimum number of driver nodes in $SC$. Given a directed network, an undirected bipartite graph is created in the following way: for every node $i$ in the original network, a node $i^+$ of type $+$ and a node $i^-$ of type $-$ are created in the bipartite graph. The connectivity in the bipartite graph is such that if node $i$ has a directed edge to node $j$ in the original network, then the bipartite graph will have an undirected edge from node $i^+$ to node $j^-$. As Liu et al. showed, a maximum matching of the bipartite graph (maximum number of edges with no common nodes) gives the minimum number of driver nodes in SC; each node in the original network corresponding to a node of type $+$ that is not in the maximum matching must be directly regulated by a driver node. A maximum matching of a graph is not unique, which implies that the set of nodes that must be directly regulated by a driver node is not unique either. The maximum matching of a bipartite graph can be efficiently found in $O(\sqrt{N}M)$ time using the Hopcroft-Karp algorithm.

For this work, we use a custom code in Python to implement the maximum matching approach of Liu et al. \cite{LiuControl}, and use the implementation of the Hopcroft-Karp algorithm in the Python package NetworkX (https://networkx.github.io/, version 1.10) to find the maximum matching.

\subsubsection*{II.D. Comparing feedback vertex set control and structural controllability}

Feedback vertex set control and structural controllability can both be used to answer the question of how difficult to control a network is, based solely on network structure, but they differ in the underlying dynamics they consider, their control objective, and their control actions. To be more specific:
\begin{enumerate}
\item[-] FC considers dissipative nonlinear dynamics, while SC considers linear dynamics or linearized nonlinear dynamics\cite{LiuControl,LiuReview}.
\item[-] FC's control objective is attractor-control (from any initial state to any target system attractor) while SC's control objective is full control (from any initial state to any target state for linear dynamics, or among states near a steady state or system trajectory for nonlinear dynamics \cite{LiuControl,LiuReview}).
\item[-] FC provides what state or trajectory the selected nodes should follow, but not how or if an external driver signal $u(t)$ (a controller) can make this happen. The existence of a driver signal $u(t)$ is guaranteed in SC, although SC does not explicitly determine this $u(t)$, and one may need to add extra constraints if $u(t)$ is required to have certain properties, e.g. the length of the linear chains of nodes spanned by each independent $u(t)$ need to be shorter than a threshold to find a numerically implementable signal \cite{SunControl,SCEnergy}.
\item[-] The external driver signal of SC is likely to be dependent on the initial state and on the target state, while the state/trajectory of the overridden nodes in FC only depends on the target attractor and is independent of the initial state (though, a controller for FC would likely be dependent on the initial condition).
\end{enumerate}
We summarize the difference between these methods in \ref{tab:STable4}. SC and FC are very different methods, so one should be careful about extending their predictions beyond their realm of applicability. Indeed, a lot of work has been done in using SC on networks which have inherently nonlinear dynamics and in which the questions asked use a notion of control that seem to be closer to attractor control (e.g. refs. \cite{PPIControl,yeastControl,brainControl}, and some of the results from refs. \cite{LiuControl,LiuReview}). In these cases, the hope seemed to be that the network insights obtained from linear dynamics would be close enough to those of nonlinear dynamics even though SC made no such guarantee. The results of our work and others' \cite{RochaControl, Bergstrom} caution against this.

\subsection*{III. Structure-based control of real networks}

\subsubsection*{III.A. Real networks used in this study}

Here we describe each network in \ref{tab:STable2}, provide the reference where each network was first reported, and give the link to where the network was obtained (if publicly available). For many of these networks, the orientation of the directed edges does not match the expected direction of influence in a dynamic model; if there is an edge from node $i$ to node $j$, we expect the state of node $i$ to influence the state of node $j$ (e.g., in an epidemic model, if individual $i$ is infected and $i$ can spread the disease to $j$, then we expect node $j$ to get infected). For these networks, we follow \cite{LiuControl} and \cite{RuthsControl}, and reverse the orientation of the directed edges in order for it to match the expected directionality of influence.

\begin{enumerate}

\item[-] \textbf{\textit{E. coli} transcription regulatory network 1 \cite{ecoli}}. Graph of the transcriptional regulation network in the bacterium \textit{Escherichia coli}. Vertices denote genes; a gene that codes for a transcription factor that regulates the transcription of a target gene is denoted by a directed edge between them. The version of the network used was obtained directly from Yang-Yu Liu.

\item[-] \textbf{\textit{E. coli} transcription regulatory network 2 \cite{ecoliAlon}}. Graph of the transcriptional regulation network in the bacterium \textit{Escherichia coli}. Operons (a gene or group of genes transcribed together) are denoted by vertices; an operon that codes for a transcription factor that directly regulates a target operon is denoted by a directed edge. This network was obtained from Hawoong Jeong's website http://stat.kaist.ac.kr/index.php.

\item[-] \textbf{\textit{S. cerevisae} transcription regulatory network 1 \cite{yeast}, 2 \cite{yeastAlon}}. Graph of the transcriptional regulation network in the yeast \textit{Saccharomyces cerevisiae}. Genes are denoted by vertices; a gene that codes for a transcription factor that regulates a target gene is denoted by a directed edge between them. Network 1 was obtained from the supplemental information in ref. \cite{yeast}, and network 2 was obtained from Uri Alon's website https://www.weizmann.ac.il/mcb/UriAlon/ download/collection-complex-networks.

\item[-] \textbf{US corporate ownership \cite{USCorps}}. Graph of the ownership relations among companies in the telecommunications and media industries in the United States. Companies are denoted by vertices and ownership of a company by another is denoted by an edge originating from the owner company. This network was obtained from the Pajek network dataset http://vlado.fmf.uni-lj.si/pub/networks/data/econ/Eva/Eva.htm

\item[-] \textbf{\textit{E. coli}, \textit{S. cerevisae}, \textit{C. elegans} metabolic networks \cite{Metabolic}}. Graph of the metabolic network of the bacterium \textit{Escherichia coli}, the yeast \textit{Saccharomyces cerevisiae}, and the worm \textit{Caenorhabtitis elegans}. Substrates (molecules) and temporary complexes are denoted by vertices; substrates that participate as a reactant in the reaction associated to a complex have an edge to it, and substrates that are products of the reaction associated to a complex have an edge from it. These network were obtained from Hawoong Jeong's website http://stat.kaist.ac.kr/index.php.

\item[-] \textbf{\textit{C. elegans} neural network \cite{Watts,Celegansneural}}. Graph of the \textit{Caenorhabtitis elegans} worm's neural network. Neurons are denoted by vertices and synapse/gap junctions between neurons are denoted by edges. This network was obtained from the UC Irvine Network Data Repository http://networkdata.ics.uci.edu/data/celegansneural/.

\item[-] \textbf{Ythan \cite{ythan,foodweb}, Seagrass \cite{seagrass,foodweb}, Grassland \cite{grassland,foodweb}, and Little Rock \cite{littlerock,foodweb} food web networks}. Graph of the predatory interactions among species in the Ythan Estuary, the St. Marks Seagrass, the England/Wales Grassland, and the Little Rock Lake. Every species is denoted by a vertex, and if a species preys on another species an edge is drawn from the prey to the predator. This network was obtained from the Cosin Project network data http://www.cosinproject.eu/extra/data/foodwebs/ WEB.html.

\item[-] \textbf{Political Blogs \cite{polblogs}}. Graph of the hyperlinks between blogs on US politics in 2005. Every blog is denoted by a vertex and hyperlinks are denoted by edges that point towards the linked blog. In this work we reverse the edges of this network so that they match the direction of influence in a dynamic model (i.e., if a blog has a hyperlink to another blog, then the latter influenced the former). This network was obtained from Mark Newman's website http://www-personal.umich.edu/$~$mejn/netdata/.

\item[-] \textbf{WWW network of stanford.edu \cite{stanfordedu} and nd.edu \cite{ndedu}}. Graph of the web networks of Stanford University (domain stanford.edu) and the University of Notre Dame (domain nd.edu). Every webpage is denoted by a vertex and hyperlinks are denoted by edges that point towards the linked webpage. This network was obtained from the Stanford Large Network Dataset Collection https://snap.stanford.edu/data/.

\item[-] \textbf{Internet networks \cite{gnutella1,gnutella2}}. Graphs of the Gnutella peer-to-peer file sharing network from August 2002; each graph represents a different snapshot of the Gnutella network. Every host is denoted by a vertex and a connection from one host to another is denoted by an edge that points towards the latter. These networks were obtained from the Stanford Large Network Dataset Collection https://snap.stanford.edu/data/.

\item[-] \textbf{Electronic Circuits \cite{yeastAlon,circuits1,circuits2}}. Network representations of electronic circuits from the ISCAS89 benchmark collection. Logic gates and flip-flops are represented by vertices, and the directed connections between them are denoted edges. These networks were obtained from Uri Alon's website https://www.weizmann.ac.il/mcb/UriAlon/ download/collection-complex-networks.

\item[-] \textbf{Texas power grid \cite{powergrid}}. Network representation of the Texas power grid. Substations, generators, and transformers are represented by vertices, and transmission lines between them are denoted by edges, with the edge directionality corresponding to the electric power flow. This network was obtained directly from Yang-Yu Liu.

\item[-] \textbf{Slashdot \cite{stanfordedu}}. Friend/foe network of the technology-related news website Slashdot obtained in 2009. Users are denoted by vertices, and a user tagging another user as a friend/foe is denoted by an edge pointing towards the latter user. In this work we reverse the edges in this network so that they match the direction of influence in a dynamic model (i.e., if a user tags another user, the latter has an influence on the former). This network was obtained from the Stanford Large Network Dataset Collection https://snap.stanford.edu/data/.

\item[-] \textbf{Wikivote \cite{wikivote1,wikivote2}}. Who-votes-for-whom network of Wikipedia users for administrator elections. Users are denoted by vertices, and a user voting for another user is denoted by an edge pointing towards the latter user. In this work we reverse the edges of this network so that they match the direction of influence in a dynamic model (i.e., if a user votes for another user, the latter has an influence on the former). This network was obtained from the Stanford Large Network Dataset Collection https://snap.stanford.edu/data/.

\item[-] \textbf{College student and prison inmate trust networks \cite{alontrust,college,prison}}. Social networks of positive sentiment of college students in a course about leadership and of inmates in prison. Each person is denoted by a vertex, and the expression of a positive sentiment of a person towards another person (based on a questionnaire) is denoted by an edge pointing towards the latter. In this work we reverse the edges of this network so that they match the direction of influence in a dynamic model (i.e., if a person has a positive sentiment towards another, the latter has an influence on the former). These networks were obtained from Uri Alon's website https://www.weizmann.ac.il/mcb/UriAlon/ download/collection-complex-networks.

\item[-] \textbf{Epinions \cite{epinions}}. Who-trusts-whom online social network of Epinions.com, a general consumer review site. Users are denoted by vertices, and a user trusting another user is denoted by an edge pointing towards the latter. In this work we reverse the edges of this network so that they match the direction of influence in a dynamic model (i.e., if a user trusts another user, the latter has an influence on the opinion of the former). This network was obtained from the Stanford Large Network Dataset Collection https://snap.stanford.edu/data/.

\item[-] \textbf{arXiv's High Energy Physics - Theory and High Energy Physics - Phenomenology citation networks \cite{arxiv1,arxiv2}}.
 Citations between preprints in the e-print repository arXiv for the High Energy Physics - Theory (hep-th) and High Energy Physics - Phenomenology (hep-ph) sections. The citations cover the period from January 1993 to April 2003. Each preprint in the network is denoted by a vertex; a preprint citing another preprint is denoted by a directed edge from the citing preprint to the cited preprint. In this work we reverse the edges of this network so that they match the direction of influence in a dynamic model (i.e., if a preprint is cited by another preprint, the latter had an influence on the former). This network was obtained from the Stanford Large Network Dataset Collection https://snap.stanford.edu/data/.

\item[-] \textbf{UC Irvine online social network \cite{uci}}. Network of messages among users in an online community for students at University of California, Irvine. Users are denoted by vertices, and a user messaging another user is denoted by an edge pointing towards the latter. This network was obtained from Tore Opsahl's website https://toreopsahl.com/datasets/.

\item[-] \textbf{Cellphone communication network \cite{cellphone}}. Call network of a subset of anonymized cellphone users. Each user is denoted by a vertex, and a call or text message from one user to another is denoted by a directed edge from the sender to the receiver. This network was obtained directly from Yang-Yu Liu.

\item[-] \textbf{E-mail communication network \cite{email}}. Network of e-mails sent among users in a university during a period of 83 days. Each user is denoted by a vertex, and an e-mail sent from one user to another during this period of time is denoted by an edge from the sender to the receiver. This network was obtained directly from Yang-Yu Liu.

\item[-] \textbf{Intra-organizational Freeman networks \cite{free}}. Network of personal relationships among researchers working on social network analysis at the beginning and at the end of the study. Each researcher is denoted by a vertex, and a personal relationship from a researcher to another is denoted by a directed edge from the former to the latter. In this work we reverse the edges of this network so that they match the direction of influence in a dynamic model (i.e., if a researcher has a personal relationship with another, the latter has an influence on the former). This network was obtained from Tore Opsahl's website https://toreopsahl.com/datasets/.

\item[-] \textbf{Intra-organizational consulting and manufacturing networks \cite{intra}}. Network describing the relationships between employees in a consulting company and in a research team from a manufacturing company. Each employee involved is denoted by a vertex, and the frequency/extent of information or advice an employee obtains from another (as measured by a questionnaire) is denoted by a weighted, directed edge among them that points from the questioned employee. We follow \cite{LiuControl} and \cite{RuthsControl}, and use all edges with a nonzero weight to define a unweighted network, which we use for our analysis. We also reverse the edges of this network so that they match the direction of influence in a dynamic model (i.e., if an employee receives advice or information from another, the latter has an influence on the former). This network was obtained from Tore Opsahl's website https://toreopsahl.com/datasets/.

\end{enumerate}

\subsubsection*{III.B. Notes on the ensembles of randomized real networks}

We study the control properties of ensembles of randomized real networks using four randomization procedures. We follow \cite{LiuControl} and \cite{RuthsControl} in using full randomization, which turns the network into a directed Erd\H{o}s-R\'{e}nyi network with $N$ nodes and $M$ edges \cite{NewmanBook}, and degree-preserving randomization, which keeps the in-degree and out-degree of every node but shuffles its successor and predecessor nodes \cite{DegRandomized}. Erd\H{o}s-R\'{e}nyi randomization is implemented by creating a graph of $N$ nodes, randomly (uniformly) choosing a source and a target of an edge from the set of $N$ nodes, and repeating this for each of the $M$ edges. For the degree-preserving randomization, we start from the original network and choose two edges randomly (uniformly), for which we switch their target nodes if the target and source nodes of both edges are each different (if they are the same, we choose another edge pair). We repeat this step for a transient of $25M$ times, after which we save the obtained network as the first element of the ensemble. We then repeat the target-node-switching step $5M$ times, save the resulting network as the second element of the ensemble, and repeat the target-node-switching step $5M$ times for each consequent ensemble element.

To verify that the cycle structure explains the observed FC node set size, we designed two new randomization procedures: an SCC-preserving and degree-preserving randomization in which the directed acyclic part of the graph is randomized and every edge that is part of an SCC is kept intact, and a short-cycle-preserving and degree-preserving randomization in which the randomized network is guaranteed to have every edge that is part of a short cycle in the original network.

For the SCC-preserving randomization, we first remove all edges that are part of an SCC, which leaves a network that is a directed acyclic graph (DAG, i.e., a graph with no cycles \cite{NewmanDAG}). Starting from this DAG, we generate a topological order $O=\{L(i)\}$ for each node $i$ in the network in the following way \footnote{In a topological order, each node $i$ is assigned a positive integer $L(i)$ in such a way that for all pairs of nodes $i$,$j$ if node $i$ has an outgoing edge to node $j$ then $L(j)<L(i)$. A topological order exists for a graph if and only if the graph is a DAG.}:
\begin{enumerate}
\item[1.-] Set $order=0$
\item[2.-] Set $L$ to be the sink nodes in the DAG.
\item[3.-] Randomly (uniformly) select a node from $L$ in the DAG, assign to it the value $L(i)=L$, and remove the selected node from the DAG.
\item[4.-] Repeat 2 and 3 with the updated DAG and increase the value $order$ by 1 at after each repeat
\end{enumerate}
The result is a topological order $O$, which we use to generate the randomized network by following the same edge-rewiring procedure as in degree-preserving randomization but only accept an edge-rewiring step if it preserves the topological order $O$. We repeat the edge-rewiring step for a transient of $25M$ times, after which we add back the edges in the SCCs and save the obtained network as an element of the ensemble. Each topological order $O$ is chosen at random to make sure that the resulting network ensemble is not generated with a single topological order like some previous work on DAGs has \cite{NewmanDAG} \footnote{We note that the algorithm we use to generate a topological order samples from every possible topological order but does not sample them uniformly. One can show that the probability $P$ of a topological order $O$ is given by $P(O)=1/C(O)$, $C(O)=l_1 \cdot l_2 \cdots l_N$, where $l_i$ is the number of elements in the list $L$ at iteration $i$ of the algorithm. Given that the objective of the algorithm is that the network ensemble is not generated with a single topological order, we consider this non-uniform sampling acceptable.}

For the short-cycle-preserving randomization, we first remove all edges that are part of a cycle of length 4 or less, and follow the same edge-rewiring procedure as in degree-preserving randomization but only accept an edge-rewiring step if does not create a cycle of length 1 or 2. We repeat this step for a transient of $25M$ times. We then do the same edge-rewiring step for a uniformly chosen edge on each cycle of length 4, a process which we repeat $10$ times. This last step is repeated but for cycles of length 3. In the resulting network we add back the short-cycle edges of the original network, and save the resulting network as the first element of the ensemble. For every other element of the ensemble, we repeat the same procedure but use a transient of $5M$ edge rewiring steps. For the short-cycle-preserving randomization of some networks, we omit the edge-rewiring step for cycles of length 4 (emails, political blogs, and UCI) or of length 3 and 4 (slashdot, wikivotes, nd.edu, Manufacturing, and epinions) because of the size and large number of cycles in these networks.

For each real network we used $\Omega=100$ networks as the ensemble size for the Erd\H{o}s-R\'{e}nyi (\ref{fig:FVSrealSI}c) and degree-preserving randomizations, and $\Omega=50$ for the SCC-preserving and short-cycle-preserving randomizations. For most ensemble properties we used the 100 ensemble networks to estimate the average value and standard deviation of the property, but for some properties this was too computationally expensive for very large networks (e.g. $FVS$ of networks with $> 2.5\times10^4$ nodes) or for very dense networks (e.g. cycle numbers of intra-organizational networks). For these properties and networks, we used a smaller ensemble size, as specified below.
\ \ \\
\ \ \\
\noindent
- Political blogs. For cycle numbers of length $4$, $\Omega=10$.\\
- nd.edu. For cycle numbers of length $4$, $\Omega=2$. For $N_{FVS}^{ER}$ and $N_{FVS}^{Rand-deg}$, $\Omega=5$ and $1$ iteration for GRASP. For $N_{FVS}^{Rand-SCC}$ and $N_{FVS}^{Rand-Cyc}$, $\Omega=50$ and $1$ iteration for GRASP.\\
- stanford.edu. For cycle numbers of length $\geq2$, $\Omega=20$. For $N_{FVS}$, $\Omega=5$ and $1$ iteration for GRASP. For the SCC-preserving and short-cycle-preserving randomization we omitted this network because of time and resource constraints.\\
- Slashdot. For cycle numbers of length $\geq3$, $\Omega=20$. For $N_{FVS}^{ER}$, $N_{FVS}^{Rand-deg}$, $N_{FVS}^{Rand-SCC}$, and $N_{FVS}^{Rand-Cyc}$, $\Omega\geq40$ and $2$ iterations for GRASP.\\
- Epinions. For $N_{FVS}^{ER}$, $N_{FVS}^{Rand-deg}$, $N_{FVS}^{Rand-SCC}$, and $N_{FVS}^{Rand-Cyc}$, $\Omega=50$ and $\geq2$ iterations for GRASP.\\
- arXiv HepTh, HepPh. For cycle numbers of length $\geq2$, $\Omega=50$. For $N_{FVS}^{ER/Rand-deg/Rand-SCC/Rand-Cyc}$, $\Omega\geq50$ and $\geq10$ iterations for GRASP.\\
- UCIonline. For cycle numbers of length $4$, $\Omega=10$.\\
- Cellphone. For $N_{FVS}^{ER}$ and $N_{FVS}^{Rand-deg}$, $\Omega=200$ and $50$ iterations for GRASP. For $N_{FVS}^{Rand-SCC}$ and $N_{FVS}^{Rand-Cyc}$, $\Omega=50$ and $25$ iterations for GRASP.\\
- Emails. For cycle numbers of length $\ge2$, $\Omega=5$.\\
- Manufacturing. For cycle numbers of length $\geq2$, $\Omega=20$.

\subsubsection*{III.C. Comparing feedback vertex set control and structural controllability in real networks}

SC was applied to diverse types of real networks and the ratio of the minimal number of SC nodes needed, $N_{SC}$, and the total number of nodes, $n_{SC}=N_{SC}/N$ was used to gauge how difficult it is to control these networks \cite{LiuControl}. Both SC and FC can be used to answer the question of which nodes need to be controlled in order to control a network (albeit they differ in the underlying dynamics they consider, their control objective, and their control actions), so a natural question is how the fraction of control nodes in real networks compares between SC and FC ($n_{FC}=N_{FC}/N$, where $N_{FC}$ is the size of the minimal FC control set). To answer this question, we apply SC and FC to the real networks in \cite{LiuControl}, and compare the fraction of control nodes $n_{SC}$ and $n_{FC}$ (\ref{fig:FVSvsSC2}a and \ref{tab:STable2}). A surprising result is that the fraction of control nodes $n_{SC}$ and $n_{FC}$ appears to be inversely related across several types of networks. For example, gene regulatory networks require between 75\% - 96\% of nodes in SC yet only require between 1\% - 18\% of nodes in FC. A similar $n_{SC}>>n_{FC}$ relationship is also seen in food web networks and internet networks, while the opposite relationship ($n_{SC}<<n_{FC}$) is seen in the social trust networks with low $n_{SC}$ and intra-organizational networks.

To explain the topological properties underlying the difference in $n_{SC}$ and $n_{FC}$, we note that the fraction of nodes $n_{SC}$ and $n_{FC}$ obey the relations
\begin{align}
n_{SC} &= n_s + n_e + n_i, \label{eq:4} \\
n_{FC} &= n_s + n_{FVS}, \label{eq:5}
\end{align}
where $n_s$ is the fraction of source nodes, $n_e$ is the fraction of external dilations nodes in SC, $n_i$ is the fraction of internal dilation nodes in SC, and $n_{FVS}$ is the fraction of nodes in the FVS of the network. Empirical directed networks tend to have a bow-tie structure \cite{Broder,NewmanBook}, in which most of the network belongs to the largest strongly connected component (which contains most cycles in the network, and thus determines $n_{FVS}$), its in-component (the nodes that can reach the strongly connected component, which thus determine $n_s$), or its out-component (the nodes that can be reached from the strongly connected component, which thus determine $n_e$). We define the fractions $\eta_x=n_x/(n_s + n_e + n_i + n_{FVS})$, where $x=s, e, i, FVS$. These fractions reflect the potential domination of a network component over the others. Eqs. \ref{eq:4}-\ref{eq:5} and the bow-tie structure of real networks offer a topological explanation for the observed relationships between $n_{SC}$ and $n_{FC}$.

Applying this reasoning to the studied real networks (\ref{tab:STable2}), we find that all networks with $n_{SC}<n_{FC}$ have a topology dominated by their SCC component ($\eta_{FVS} >> \eta_{e}, \eta_{i}, \eta_{s}$, \ref{fig:FVSvsSC2}, brown shading; e.g. intra-organizational networks, the college students and prison inmates trust networks, and the \textit{C. elegans} neural network). Most networks with $n_{SC}>n_{FC}$ are dominated by their out-component ($\eta_e >> \eta_{FVS}, \eta_{i}, \eta_{s}$, \ref{fig:FVSvsSC2}, yellow shading; e.g. gene regulatory networks, most food webs, and internet networks) or by internal dilations ($\eta_i >> \eta_{FVS}, \eta_{e}, \eta_{s}$, \ref{fig:FVSvsSC2}, pink shading; e.g. metabolic networks and circuits). The rest of the networks have a mixed profile ($\eta_{FVS} \simeq \eta_{e} \simeq \eta_i \simeq \eta_{s}$, \ref{fig:FVSvsSC2}, no shading), and include networks with $n_{SC}>n_{FC}$ (citation networks and the Texas power grid) and the networks in which $n_{SC}\simeq n_{FC}$ (a political blog network and two online social communication networks).

\subsection*{IV. Structure-based control of the \textit{Drosophila melanogaster} segment polarity gene regulatory network}

We compare the results of the two control methods for the gene regulatory network of the Drosophila segment polarity genes, for which several dynamic models exist \cite{SegPolODE,SegPolBool,Chaves}. The segment polarity genes, especially wingless (\textit{wg}) and engrailed (\textit{en}), are important determinants of embryonic pattern formation and contributors to embryonic development \cite{SegPolODE}. The wingless mRNA and protein are expressed in the cell that is anterior to the cell that expresses the engrailed and hedgehog (\textit{hh}) mRNA and protein. All models consider a group of four subsequent cells as a repeating unit, and include intra-cellular and inter-cellular interactions.

The continuous model of von Dassow et al. represents each cell as a hexagon with six relevant cell-to-cell boundaries. It includes 136 nodes that represent mRNAs and proteins, among them 4 source nodes and 24 sink nodes, and 488 edges that represent transcriptional regulation, translation, and protein-protein interactions. Fig. 4a in the main text, reproduced here as \ref{fig:SFig2}a, shows the network corresponding to the \textit{wg}-expressing cell (cell 1) and three of its boundaries with the \textit{en}-expressing cell 2. Additional nodes in the network include, \textit{ptc} (patched), \textit{ci} (cubitus interruptus), its proteins \textit{CID} and \textit{CN} (repressor fragment of \textit{CID}), \textit{IWG} (intracellular \textit{WG} protein), \textit{EWG} (extracellular \textit{WG} protein), \textit{PH} (complex of patched and hedgehog proteins), and \textit{B}, a constitutive activator of \textit{ci}. For each gene, the mRNA is written in lower case and the protein(s) are written in upper case. The nodes are characterized by continuous concentrations, whose rate of change is described by ordinary differential equations (ODE) involving Hill functions for gene regulation and mass action kinetics for protein-level processes, and using 48 kinetic parameters \cite{VonDassow1,VonDassow2}. von Dassow et al. have shown that the model can reproduce the essential feature of the wild type steady state: \textit{wg}/\textit{WG} are expressed anterior to the parasegment boundary (cell 1) and \textit{en}/\textit{EN}/\textit{hh}/\textit{HH} are expressed posterior to the parasegment boundary (cell 2) as shown in Fig. 4. The initial condition that yields this steady state for the most parameter sets, the so-called `` crisp'' initial condition, \textit{wg}/\textit{IWG} in the first cell is at maximal concentration (1), \textit{en}/\textit{EN} in the second cell has concentration 1, the source nodes B are fixed at 0.4 in each cell and all the other nodes have zero concentration.

Wild type steady state of the von Dassow et al. model for the second parameter set provided by the \textit{Ingeneue} program \cite{VonDassow1,VonDassow3}, using normalized concentration variables
\begin{align*}
c(en_2) &= c(EN_2)= 0.986, \\
c(wg_1) &= 0.857, \\
c(IWG_1) &= 0.006, \\
c(EWG_{0,0}) &= c(EWG_{0,3-5}) = 0.005,\\
c(EWG_{0,1}) &= c(EWG_{0,2}) = 0.011, \\
c(EWG_{1,0})&= c(EWG_{1,3}) =0.269, \\
c(EWG_{1,1-2}) &= c(EWG_{1,4-5}) = 0.264,\\
c(EWG_{2,0-3}) &= 0.005,\\
c(EWG_{2,4}) &= c(EWG_{2,5}) = 0.011,\\
c(ptc_0) &= c(ptc_1) = c(ptc_3)=0.995,\\
c(ptc_2) &= 0.001,\\
c(PTC_{0,*}) &= c(PTC_{1,*}) = c(PTC_{3,*})=0.166,\\
c(ci_0) &= c(ci_1) = c(ci_3)=0.868,\\
c(ci_2) &=0.007,
\end{align*}
\begin{align*}
c(CI_0) &= c(CI_1) = c(CI_3)=0.057,\\
c(CI_2) &=0.005,\\
c(CN_0) &= c(CN_1) = c(CN_3)=0.42,\\
c(CN_2) &=0.001,\\
c(hh_2) &= 1,\\
c(HH_{2,0}) &= c(HH_{2,3})=0.072,\\
c(PH_{1,1-2}) &= c(PH_{3,4-5}) = 0.001,
\end{align*}
where $i,*$ represents all sides of the $i$th cell. The concentration of the other nodes is smaller than $10^{-5}$.

Another initial condition considered here is a nearly-null initial condition, wherein intra-cellular nodes have a concentration of 0.05 in the first and third cell and 0.15 in the second and fourth (zeroth) cell; membrane-localized nodes have concentration of 0.15 for even-numbered sides and 0.05 for odd-numbered sides in every cell. This initial condition yields an unpatterned steady state for the majority of parameter sets.

Unpatterned steady state of the von Dassow et al. model, for the second parameter set provided by the \textit{Ingeneue} program \cite{VonDassow1,VonDassow3}, using normalized concentrations:
\begin{align*}
c(wg_*)&=0.857, \\
c(IWG_*)&=0.007 \\
c(EWG_{*,*})&=0.28,\\
c(ptc_*)&=0.996,\\
c(PTC_{*,*})&=0.166,\\
c(ci_*)&=0.868,\\
c(CI_*)&= 0.057,\\
c(CN_*)&= 0.42,
\end{align*}
where $*$ represents for all cells, and $*,*$ represents for all sides in all cells. The concentration of the other nodes is smaller than $10^{-5}$.

The differential equation system is solved using a custom code in Python and the odeint function with default parameter setting. We used the differential equations given in the appendix of \cite{VonDassow2}. \textit{Ingeneue} can be found at http://rusty.fhl.washington.edu/ingeneue/papers/ papers.html.

The Boolean model implements a few modifications in the network topology compared with the ODE network model, and considers only two cell-to-cell boundaries instead of six. There are 56 nodes and 144 edges in the network as shown in Fig. 4b. One difference compared with the von Dassow et al. model is the existence of three cubitus interruptus proteins: the main protein \textit{CI}, and two derivatives with opposite function: \textit{CIA}, which is a transcriptional activator, and \textit{CIR}, a transcriptional repressor. There are four source nodes, representing the sloppy paired protein (\textit{SLP}), which is known to have a sustained expression in two adjacent cells (cells 0 and 1 if the \textit{wg}-expressing cell is considered cell 1) and is absent from the other two. There are ten steady states for this Boolean network model when considering the biologically relevant pattern of the source node states. Starting from the biologically known wild type initial condition, which consists of the expression (ON state) of $SLP_0$, $SLP_1$, $wg_1$, $en_2$, $hh_2$, $ci_0$, $ci_1$, $ci_3$, $ptc_0$, $ptc_1$, $ptc_3$, the model converges into the biologically known wild type steady state illustrated on Fig. 4c.

Specifically, the wild type steady state of the Albert \& Othmer model consists of the expression of
\begin{align*}
SLP_0, SLP_1, wg_1, WG_1, en_2, EN_2, hh_2, HH_2,\\
ci_0, ci_1, ci_3, CI_0, CI_1, CI_3, CIA_1, CIA_3, CIR_0,\\
ptc_1, ptc_3, PTC_0, PTC_1, PTC_3, PH_1, PH_3.
\end{align*}

Analytical solution reported in \cite{VonDassow2} indicated that the states of the \textit{wg} and \textit{PTC} nodes, each of which has a positive auto-regulatory loop, determine the steady state for the given source node (\textit{SLP}) configuration \cite{SegPolBool}. For example, any initial condition with no \textit{wg} expression leads to an unpatterned steady state wherein \textit{ptc}, \textit{ci}, \textit{CI} and \textit{CIR} are expressed in each cell, and the rest of the nodes are not expressed in any cell.

\subsubsection*{IV.A. Structure-based control of the von Dassow et al. differential equation model} \

The FC method predicts that one needs to control $N_{FC}=52$ nodes (4 source nodes and 48 additional nodes) to lead any initial condition to converge to any original attractor of the model. There are multiple control sets with $N_{FC}=52$; one of them consists of \textit{B} (source node), \textit{CI}, \textit{CN}, \textit{IWG}, \textit{EWG} on every other side, \textit{HH} on every other side, \textit{PTC} on every other side in all four cells (shown in \ref{fig:SFig2}a). We perform simulations using two benchmark parameter sets to test this prediction. We use the second parameter set provided by the Ingeneue program to test the system's convergence to a steady state \cite{VonDassow1,VonDassow3}. The ODE system has at least two steady states with this parameter set. A nearly null initial condition leads to the unpatterned state (illustrated by the green lines in Fig. 4d in the main text). The crisp initial condition leads to the wild type pattern (see pink lines in Fig. 4d), which we choose as the desired steady state. If we start from the nearly null initial condition and maintain the concentrations of the nodes in the FC node set in the values they would have in the desired steady state, the system evolves into the desired steady state (see blue lines and inset of Fig. 4d). We obtained the same success of FC control when starting from 100 different random initial conditions (shown in \ref{fig:SFig3}a). We also obtained the same success using a reduced FC set (blue lines in \ref{fig:SFig3}b), which consists of \textit{B}, \textit{CID}, \textit{CN}, \textit{IWG} in every cell. In contrast, in the absence of control none of the trajectories converge to the wild type steady state (red lines in \ref{fig:SFig3}b).

We also numerically verified, using a different benchmark parameter set, namely the first parameter set provided by the Ingenue program, that FC control can also successfully drive any state to a limit cycle attractor (see \ref{fig:SFig4}a). This limit cycle attractor has the same expression pattern of \textit{en}, \textit{wg} and \textit{hh} as the wild type steady state, thus we refer to it as the wild type limit cycle (illustrated in \ref{fig:SFig4}c). We also obtained the same success of driving any state to a limit cycle attractor using the same reduced Feedback vertex control shown in \ref{fig:SFig4}b.

SC control indicates multiple control sets with $N_{SC}=24$ nodes. One possible combination is $B_*$, $PTC_{*,1}$, $PTC_{*,3}$, $PTC_{*,5}$, $HH_{*,5}$, $PH_{*,1}$, where $*$ represents all cells (shown in \ref{fig:SFig2}b. Though SC predicts that less nodes need to be controlled, applying it requires a potentially complicated time-varying driver signal, which would need to be determined for each initial condition using, for example, minimum-energy control or optimal control \cite{LiuReview,OptimalControl}.

\subsubsection*{IV.B. Structure-based control of the Albert \& Othmer Boolean model}

The FC method predicts that $N_{FC}=14$ nodes need to be controlled, including the 4 source nodes (\textit{SLP}), the 8 self-sustaining nodes (all \textit{wg} and \textit{PTC}), and 2 additional nodes (with one possibility being $CIR_1$ and $CIR_3$). Since the FC set contains all \textit{wg} and PTC nodes, which were shown to determine the steady states under the indicated source node states, we can conclude that controlling the nodes in the FC set is enough to drive any initial condition to the desired steady state in the Albert \& Othmer model. The simulation result is consistent with the theoretical result, as shown in Fig. 4e. The wild type initial condition leads to the wild type steady state (pink lines). The null initial condition used in the Boolean model is that all the nodes are in the OFF state; the resulting steady state is the unpatterned steady state (green lines). The controlled trajectory with FC is shown in blue lines. We obtained the same success of FC control when starting from 100 different random initial conditions, as shown in \ref{fig:SFig5}a. Moreover, the 12 nodes consisting of \textit{SLP}, \textit{wg} and \textit{PTC} in each cell (which we refer to as the reduced FC set) are enough to drive all the random initial conditions to the desired steady state in this particular model, as shown in \ref{fig:SFig5}b.

SC control predicts that we only need to control the four source nodes (\textit{SLP}), as the network can be covered by four branches and one loop. Relevant to this, Albert \& Othmer studied three scenarios of fixed states of the source nodes. If the source nodes are locked into their respective states in the wild type steady state (two ON and two OFF), there are six reachable attractors, one of which is the wild type steady state. If all source nodes are locked into the OFF state, there are seven attractors, but none of them is the wild type steady state. If all source nodes are locked into the ON state, the unpatterned state is the only attractor. These results suggest that the correct expression of the source nodes is necessary, but not sufficient for attractor control of the system. Indeed, SC can make no such guarantee, since for general nonlinear systems it only provides sufficient conditions for local controllability around a steady state or a system trajectory.

For a simplified, single-cell version of the Albert \& Othmer model, Gates and Rocha showed that the SC node set is sufficient for attractor control, but does not fully control this system \cite{RochaControl}. Thus, a control method such as \cite{AkutsuBool,ChengBool} seems to be required for correctly predicting full control node sets in Boolean models.

\end{document}